\documentclass[preprint,11pt]{elsarticle}

\usepackage{graphicx}
\usepackage{lscape}
\usepackage{upgreek}
\usepackage{caption} 
\usepackage{hyperref}
\usepackage{todonotes}
\usepackage{amssymb}
\usepackage{lineno}
\usepackage{float}

\usepackage{color,soul}

\journal{Journal of System and Software}

\usepackage{longtable}
\usepackage{multirow}
\usepackage{soul}

\newcommand{\quickwordcount}{
  \immediate\write18{texcount -1 -sum -merge \jobname.tex > \jobname-words.sum }
  \input{\jobname-words.sum} words
}

\begin{document}

\begin{frontmatter}

\title{Are Architectural Smells Independent from Code Smells? An Empirical Study}

\author[myfirstaddress]{Francesca Arcelli Fontana}
\ead{francesca.arcelli@unimib.it}

\author[mysecondaddress]{Valentina Lenarduzzi}
\ead{valentina.lenarduzzi@tuni.fi}

\author[mythirdaddress]{Riccardo Roveda}
\ead{riccardo.roveda@alten.it}

\author[mysecondaddress]{Davide Taibi}
\ead{davide.taibi@tuni.fi}

\address[myfirstaddress]{University of Milano-Bicocca, Milan (Italy)}

\address[mysecondaddress]{Tampere University, Tampere (Finland)}

\address[mythirdaddress]{Alten Italia, Milano, (Italy)}

\begin{abstract}
\textit{Background.} Architectural smells and code smells are symptoms of bad code or design that can cause  different quality problems, such as faults, technical debt, or difficulties with maintenance and evolution.
Some studies show that code smells and architectural smells often appear together in the same file. The correlation between code smells and architectural smells, however, is not clear yet; some studies on a limited set of projects have claimed that architectural smells can be derived from code smells, while other studies claim the opposite. \\
\textit{Objective.} The goal of this work is to understand whether architectural smells are independent from code smells or can be derived from a code smell or from one category of them.  \\
\textit{Method.} We conducted a case study analyzing the correlations among 19 code smells, six categories of code smells, and four architectural smells. \\
\textit{Results.} The results show that architectural smells are correlated with code smells only in  a very low number of occurrences  and therefore cannot be derived from code smells.\\
\textit{Conclusion.} Architectural smells are independent from code smells, and therefore deserve special attention by researchers, who should investigate their actual harmfulness, and practitioners, who should consider  whether and when to remove them.  
\end{abstract}


\end{frontmatter}


\section{Introduction}
\label{sec:Introduction}

Architectural smells, as introduced by Garcia et al.~\cite{Garcia2009}, are  ''Architectural decisions that negatively impact system internal quality.
Architectural smells may be caused by applying a design solution in an inappropriate context, mixing design fragments that have undesirable  emergent behaviors, or applying design abstractions at the wrong level of granularity.'' 
Several studies claim that architectural smells lead to architectural erosion and that architectural issues are the greatest source of technical debt~\cite{Ernst2015}, \cite{Nord2012}. Hence, they have to be considered as one of the primary sources of investigation for mitigating the problem of architecture degradation~\cite{Nord2012}.
The code infected by an architectural smell is a natural candidate for refactoring in order to prevent the occurrence of critical quality issues.

Code smells 
were introduced by Fowler~\cite{Fowler1999} to describe a code structure that is likely to cause problems and that can be removed through refactoring.  They commonly increase the software's defectiveness~\cite{Palomba2018}, \cite{Hall2014} and change proneness~\cite{Palomba2018}, \cite{Olbrich2009} and increase maintenance effort~\cite{Sjoberg2013}, \cite{Lozano20008}. Unlike architectural smells, they
are defined at a lower  level of granularity and do not take into account the software architecture of the system under development, commonly focusing on class or method levels. 

Several studies have investigated the interrelations between code smells, e.g., whether a code smell leads to another code smell, or whether some code smells tend to go together~\cite{Pietrzak2006}, \cite{Bartek2018}, \cite{LIU2012}, \cite{Yamashita2015-icsme}. 
Other studies have considered the  possible correlations and the impact of code smells on various software qualities such as  defects, bugs, changes, and code understandability (\cite{Li2007},~\cite{palomba2019}, ~\cite{Deligiannis2003}, ~\cite{Bois2006}, ~\cite{Olbrich2009}). 

To the best of our knowledge, only few studies have been published that analyze the correlations between code smells and architectural smells. 
Among them, one work~\cite{Macia2012b} identified correlations between code smells and architectural smells, while  another work~\cite{Macia2012} claims that they are not correlated.
In any case, no extended empirical evaluations have been carried out and no code smell stands out as the best indicator of harmfulness with respect to architecture degradation. 

The goal of this work is to understand whether architectural smells can be derived from a  code smell, or from one category of code smells (we considered the categories proposed by Mantyla ~\cite{Mantyla2006}).
For this purpose, we designed and conducted a large empirical study on possible correlations existing between  four architectural smells, 19 code smells, and six categories of code smells, analyzing  111 Java projects taken from the Qualitas Corpus Repository~\cite{Tempero2010}. 
We considered a large set of code smells defined in the literature and four architectural smells based on dependency issues that can have a critical impact on the software quality of a project  and its progressive architecture degradation \cite{ArcelliFontana2016c}.

Hence, this study aims to assess the existence of any correlation between code smells and higher-level architectural smells. We did not consider correlations between defects and code smells, which have already been studied to a large extent in the literature, but only possible correlations between code smells and architectural smells.
The results of this work will support researchers and practitioners in understanding whether they should detect both architectural smells and code smells or whether the detection of code smells alone is enough to highlight the same anomalies that could be highlighted by an architectural smell. 
If we could find some kind of correlation between architectural smells and lower-level code smells, we could mark architecturally problematic parts of software systems for extra attention by using existing code smell detectors. More importantly, this might enable us to solve some of the higher-level problems using smaller refactorings, which would be more desirable for maintainers.
The results we obtained do not reveal any significant correlation, suggesting that architectural smells cannot be derived from code smells and practitioners should take extra care to deal with architectural smells. They cannot focus only on the refactoring of code smells, but need to pay particular attention to the more dangerous architectural smells as well.
Hence, in most of cases, code smells will infect different classes than those infected by architectural smells, which not only highlights different problems but also  different candidate classes for refactoring.


The main contributions of our work can be summarized as follows:
\begin{itemize}

\item We investigated whether architectural smells are correlated with code smells or with a specific category of code smells. Other studies have considered correlations with more general architectural problems (\cite{Macia2012b},~\cite{Macia2012}).
\item We considered a huge number of analyzed projects (111). To the best of our knowledge, previous studies  investigated smaller sets of Java projects (see Related Work section).
\item Possible correlations between code and architectural issues have not been widely explored. Existing results are contradictory - evidence that this topic deserves careful attention. Hence, through our study we further emphasize how important it is for developers and maintainers to take into account both code and architectural smells during their refactoring activities. 

\end{itemize}

\textbf{Structure of the paper}. This paper is structured as follows: Section~\ref{sec:relWork} describes some related work done by researchers in the last years, while Section~\ref{sec:Background} describes the background on which our paper is based. In  Section~\ref{sec:CaseStudy}, we present the case study, where we define the research questions, the metrics, the  hypothesis and the study context with our Research Questions as well as the data collection and  data analysis procedure. In Section~\ref{sec:Results}, we show the results obtained and discuss them in Section~\ref{sec:Discussion}.  Section~\ref{sec:Threats} focuses on threats to the validity of our study. In Section~\ref{sec:Conclusions and Future Developments}, we draw conclusions and outline some possible future work.

\section{Related Work}
\label{sec:relWork}
Many  studies on code smells can be found in the literature; they consider  different aspects such as the relationships among code smells and their impact  on different features such as faults~\cite{Li2007}~\cite{Hall2014}, maintainability~\cite{Deligiannis2003}~\cite{Kapser2008}, comprehensibility~\cite{Bois2006}, change frequency~\cite{Olbrich2009}~\cite{Schumacher2010}, change size~\cite{Olbrich2009}~\cite{Olbrich2010}, and maintenance effort~\cite{Lozano20008}~\cite{Sjoberg2013}. 
Moreover, several commercial and research tools for code smell detection have been developed~\cite{ArcelliFontana2011c} (e.g., HIST~\cite{Palomba2015}, JCodeodor~\cite{ArcelliFontana2015h}, Wekanose~\cite{Azadi2018}, JDeodorant~\cite{FokaefsTC07}).
Less work is available on architectural smells.
Hence, in this section we will describe some work found in the literature on architectural smell definitions, on code smell correlations, as well as studies considering both  code smells and architectural smells.

We need to point out that in the literature, different terms are often used to describe the same concept:
e.g., in some cases code smells are also called
 code anomalies, design flaws, design smells, design disharmonies, or antipatterns.
This is the case, for example, for the God Class design disharmony~\cite{Lanza2006}, which is similar to the Large Class code smell defined by
Fowler~\cite{Fowler1999} or to the Blob antipattern~\cite{Brown1998}.  Cyclic Dependency is called an architectural smell~\cite{Lippert2006}, but corresponds to the Tangle antipattern~\cite{ArcelliFontana2011e} and to the Cyclically-Dependent Modularization design smell ~\cite{Suryanarayana2015}, or is considered an architectural violation~\cite{Sonargraph}.

Our paper focuses particularly on possible correlations existing between code smells and architectural smells.

\subsection{Code Smell Correlations} \label{sec:background_cs_correlations}

Pietrzak and Walter~\cite{Pietrzak2006} describe several types of inter-smell correlations to support more accurate code smell detection
and to better understand the effects caused by interactions
between smells.
They found different kinds of correlations among six different code smells by analyzing the Apache Tomcat project.


Arcelli Fontana et al.~\cite{Arcelli2015-sam} analyzed 74 projects of the Qualitas Corpus, detecting six smells, some correlations among smells, and possible co-occurrence of smells.
They found a high number of correlations among  God Class and Data Class, as well as among other code smells that tend to go together, and a high number of co-occurrences of the Brain Method smell with Dispersed Coupling and Message Chains.

Liu et al.~\cite{LIU2012} propose a detection and resolution sequence  for different smells by analyzing certain code smell correlations given by commonly occurring bad smells. 
 They analyzed whether it is better to  first identify smell \texttt{A} than smell \texttt{B}, e.g., Large Class versus Feature Envy or versus Primitive Obsession, or Useless Class versus other smells. 
 They considered nine code smells and identified fifteen correlations of this kind.
 
Yamashita et al.~\cite{Yamashita2015-icsme} studied possible correlations among smells. 
They  incorporated dependency analysis in order to identify a wider range of inter-smell correlations, and analyzed one industrial and two  open-source projects. 
They found the following correlations: collocated smells among God Class, Feature Envy, and Intensive Coupling, and coupled smells between  Data Class and Feature Envy.

Moreover, various authors provide code smell classifications or taxonomies that are useful for capturing possible correlations among smells.

M\"{a}ntyl\"{a} et al.~\cite{MantyllaICSM2003} categorized all of Fowler's code smells except for Incomplete Library Class and Comments smells into five categories: 
Bloaters, Object Orientation Abusers, Change Preventers, Dispensables, Encapsulators, and Couplers. The study outlines the existence of several correlations among smells belonging to the same
 category. 

Moha et al.~\cite{Moha2007a} propose a taxonomy of smells and
describe some correlations among design smells, such as Blob and (many) Data Class, or Blob and (Large Class and Low Cohesion).
 
Lanza and Marinescu~\cite{Lanza2006} propose a classification of twelve smells, called ''design disharmonies'',
into three categories: Identity, Collaboration, and Classification
disharmonies. They describe the most common correlations between the disharmonies 
in a type of diagram called a correlation web. However, these correlations were not empirically validated.

\subsection{Architectural Smells}\label{sec:background_as}
In this section, we provide a description  of some of the architectural smells (AS) defined in the literature. 
In most studies, they are actually called architectural smells, but in a few cases they are called design smells~\cite{Suryanarayana2014} or antipatterns~\cite{Brown1998}.

Garcia et al. \cite{Garcia2009} define the Connector Envy, Scattered
Functionality, Ambiguous Interface, and Extraneous Connector AS.
They provide a description of each AS, outlining the quality impact and the trade-offs and providing a generic schematic view of each smell captured in one or more UML diagrams. 
They assert that  architects can manually use such diagrams to inspect their own designs to look for architectural smells. 

Macia~\cite{Macia2013Phd} analyzed different architectural
smells related to dependency and interface issues:
Ambiguous Interface, Redundant Interface, Overused Interface, Extraneous Connector, Connector Envy, Cyclic Dependency, Scattered Parasitic Functionality, and Component Concern Overload (Component Responsibility Overload).

Mo et al.~\cite{MoWicsa2015} and Kazman et al.~\cite{KazmanIcse2015} defined five AS, four at the file level and one at the package level, which they call Hotspot Patterns: Unstable Interface, Implicit Cross-Module Dependency, Unhealthy Inheritance Hierarchy, Cross-Module Cycle, and Cross-Package Cycle.
These AS were defined in the context of the authors' research on Design Rule Spaces (DRSpaces)~\cite{XiaoFSE2014}.
The authors also developed a tool called Hotspot Detector, which is able to detect the five AS mentioned above.
The detector takes as input  several files produced by another tool called  Titan~\cite{XiaoFSE2014}.
Currently, Hotspot Detector is being evolved into a new commercial tool.

Marinescu~\cite{Marinescu2012} defined three AS:
Cyclic Dependency, Stable Abstraction Breaker, and Unstable Dependency.
They developed a tool called inFusion, which was able to detect these architectural smells and a large number of code smells. However, this tool is no longer available.

Lippert and Rook~\cite{Lippert2006} defined different  AS at different levels by essentially considering dependency and inheritance issues and aspects related to small/large size in terms of  number of packages, subsystems, and layers.
In particular, they defined  AS  in dependency graphs,   inheritance hierarchies,  packages,  subsystems, and layers.

 Le et al. \cite{ARCADE} developed a tool for the detection of some AS
and proposed a classification of the AS based on four categories: Interface, Change, Dependency and Concern-based smells. 

Suryanarayana et al.~\cite{Ganesh2013,Suryanarayana2015} adopted an
approach for classifying and cataloging a number of recurring structural design smells based on how they violate key object-oriented design principles.
Their definition of design smells is similar to the one of architectural smells, but many of their design smells correspond to the code smells of Fowler.
They identified the following design smell categories: Abstraction, Encapsulation, Modularization, and Hierarchy. They developed a tool, called Designite, to detect different design smells in C$\#$  projects.

As we can see, different AS definitions have been proposed, but few detection tools are freely available \cite{Azadi2019}.

\subsection{Code Smells and Architectural Degradation}\label{sec:background_cs_arch_degradation}
There is little  knowledge, as outlined by Macia~\cite{Macia2013Phd}, about the extent to which code anomalies are related to architectural degradation.
In the following, we report on some studies where the term code anomalies is sometimes used instead of the term code smells and  architectural anomalies correspond to architectural smells.



Macia et al.~\cite{Macia2012} analyzed code anomaly occurrences in 38 versions of five applications using existing detection strategies.
The outcome of their evaluation suggests that many of the code anomalies detected  were not related to architectural problems.
Even worse, over 50\% of the anomalies not observed by the employed techniques (false negatives) were found to be correlated with architectural problems.

In another work, Macia et al.~\cite{Macia2012b} studied the correlations between code anomalies and architectural smells in six software projects (40 versions).
They considered five architectural smells and nine code smells.
They empirically found that each architectural problem represented by each AS is often refined by multiple code anomalies.
More than 80\% of architectural problems were found to be correlated with code anomalies.
They also found \textit{1)} that certain types of code smells, such as Long Method or God Class, were consistently correlated with architectural problems; \textit{2)} that the highest percentages of code smells that introduce architectural problems occurred for God Class, Long Method, and Inappropriate Intimacy instances, and \textit{3)} that the occurrence of both  God Class and Divergent Change smells in the same code element was a strong indicator of architectural problems, such as Scattered Functionalities violating the Separation of Concerns design principle.
However, the study revealed that no type of code smell stands out as the best indicator of harmfulness with respect to architecture degradation.

Oizumi et al.~\cite{Oizumi2016} propose studying and assessing the extent to which code smell agglomerations help developers to locate and prioritize design problems. 
They  propose  considering not only the syntactic  relations among code smells, but also the semantic relations to find more powerful smell agglomerations in order to identify design problems.
Their findings show that 50\% of syntactic agglomerations  and  80\% of semantic agglomerations are related to design problems.

Oizumi et al.~\cite{Oizumi2014} analyzed seven projects and demonstrated that  agglomerations are
better than single anomaly instances to indicate the presence of an architectural problem.
They considered six code smells detected using the rules of Lanza-Marinescu~\cite{Lanza2006} and seven architectural smells detected using the rules defined by Macia~\cite{Macia2013Phd}.

Guimaraes et al \cite{Guimares2014} conducted a
controlled experiment utilizing architecture blueprints to prioritize
various types of code smells
 and provide an analysis of whether and to what extent the use of blueprints impacts the time required for revealing architecturally relevant code anomalies.




Unlike the previous studies, 
 we \textit{1)} analyzed a total of 111 Java projects, 
\textit{2)} employed two available and validated tools to detect  code and architectural smells;
\textit{3)} analyzed 19 code smells and four architectural smells, and
\textit{4)} applied different correlation analyses.
Moreover, as previous papers did not make it clear, respectively provided not much empirical validation, whether some kind of correlation exists between code smells and architectural smells, our study is intended to provide a further investigation in this direction.


\section{Background}
\label{sec:Background}
In this Section, we present the code smells together with their proposed classification and the architectural smells adopted in this work.

\subsection{Code Smells}
\label{sec:Background_smells}
In this work, we consider code smells detected by SonarQube~\footnote{SonarQube https://www.sonarqube.org/} using the ''Antipatterns-CodeSmell'' plugin~\footnote{SonarQube https://github.com/davidetaibi/sonarqube-anti-patterns-code-smells}. All the code smells, except for Duplicated Code, are detected by the ''Antipatterns-CodeSmell''  plugin, while Duplicated Code is detected natively by SonarQube. 
Here is the list of code smells considered in this work:
\begin{itemize}
\item \emph{Anti-Singleton (ASG)}: A class that provides mutable class variables exhibiting the properties of global variables~\cite{Khomh2012}.

\item \emph{Base Class Knows Derived Class (BCKD)}: A class that does not respect the heuristic defined by Riel~\cite{Riel1996}, which says that ''Derived classes must have knowledge of their base class by definition, but base classes should not know anything about their derived classes.''~\cite{Kpodjedo2011}.

\item \emph{Base Class Should Be Abstract (BCSA)}: An inheritance tree contains roots that are not abstract - only the leaves should be concrete~\cite{DECOR}.

\item \emph{Blob (BL)}: 
The majority of the  responsibilities are allocated to a single class that monopolizes the processing. A Blob class is characterized by a class diagram composed of a single complex controller class surrounded by simple data classes.
~\cite{Brown1998}.

\item \emph{Class Data Should Be Private (DsP)}: A class that publicly exposes its variables~\cite{Taba2013}.

\item \emph{Complex Class (CC)}: A class with high MC-Cabe’s cyclomatic complexity~\cite{Marinescu2004}.

\item Duplicated Code (DC): A class or method that contains an identical piece of code of another class or method. Note that we only consider internal project duplication and not cross-project duplication. 

\item \emph{Functional Decomposition (FD)}: Non-object-oriented design (possibly from legacy) is coded in an object-oriented language and notation~\cite{Brown1998}.

\item \emph{Large Class (LC)}: A class with too many lines of code, methods, or variables~\cite{Fowler1999}.

\item \emph{Lazy Class (LzC)}: ''A class that is not doing enough to pay for itself.'' ~\cite{Fowler1999}.

\item \emph{Long Method (LM)}: A method with too many lines of code~\cite{Fowler1999}.

\item \emph{Long Parameter List (LPL)}: A method having too many parameters~\cite{Fowler1999}.

\item \emph{Many Field Attributes But Not Complex (MFnC)}: A class that is not complex but has many public fields~\cite{DECOR}.

\item \emph{Message Chains (MC)}: A chain of methods that ask for an object, which asks for another one, which asks for yet another, and so on~\cite{Fowler1999}.

\item \emph{Refused Parent Bequest (RPB)}: The subclass uses only a few features of the parent class~\cite{Fowler1999}.

\item \emph{Spaghetti Code (SC)}: An ad-hoc software structure that makes it difficult to extend and optimize the code~\cite{Brown1998}.

\item \emph{Speculative Generality (SG)}: Hooks and special cases in the code that handle things that are not required, but are speculated to be required someday~\cite{Fowler1999}.

\item \emph{Swiss Army Knife (SAK)}: Over-design of interfaces results in objects with numerous methods that attempt to anticipate every possible need. This leads to designs that are difficult to comprehend, utilize, and debug, as well as to implementation dependencies~\cite{Brown1998}.

\item \emph{Tradition Breaker (TB)}: An inherited class provides a large set of new services that are unrelated to those provided by the base class~\cite{Marinescu2004}.
\end{itemize}

\subsection{Categories  of Code Smells}
\label{sec:background_categoryOfSmells}
The categories of code smells we considered are based on the classification proposed by M\"{a}ntyl\"{a} and Lassenius \cite{Mantyla2006},
where the smells are classified according to  some of the common concepts shared by the smells within one category. 
Below, we provide a description of each category and the smells included by the authors that we were able to detect with the Antipatterns-CodeSmell tool, as well as the new smells we included in the categories, if any.

\begin{itemize}
\item \emph{The Bloaters (Bloat.)}: Objects that have grown too much and can become hard to  manage. 
This category includes the code smells \textit{Blob}, \textit{Long Method}, \textit{Large Class}, and \textit{Long Parameter List}. 
We additionally included \textit{Complex Class} and \textit{Swiss Army Knife}. 

\item \emph{The Dispensables (Disp.)}: Unnecessary code fragments that should be removed. This includes the code smells \textit{Lazy Class}, \textit{Duplicated Code}, and \textit{Speculative Generality}. 
We also included \textit{ Many Field Attributes But Not Complex}.

\item \emph{The Encapsulators (Enc.)}: Objects that present high coupling (this category is also called \emph{Couplers}).
This category includes the code smell \textit{Message Chain}.

\item \emph{The Object-Orientation Abusers (OOA)}: Classes that do not comply with object-oriented design.
For example, a Switch Statement, even if applicable in procedural programming, is highly deprecated in object-oriented programming. This category includes the code smells \textit{Anti-Singleton} and \textit{Refused Parent Bequest}. 
We also included \textit{Base Class Knows Derived Class}, \textit{Base Class Should Be Abstract}, \textit{Class Data Should Be Private}, and \textit{Tradition Breaker}. 

\item  \emph{The Change Preventers}: This category includes smells that hinder further changes in the source code.
This category includes a set of code smells such as \textit{Divergent Change}, \textit{Shotgun Surgery}, and \textit{Parallel Inheritance Hierarchies}, which are not detected by the Antipatterns-CodeSmell tool.
We also included \textit{Spaghetti Code}.
\end{itemize}

Moreover, since we believe that some code smells considered in this work could be grouped together, we defined a new category:
\begin{itemize}
\item \emph{The Object-Oriented Avoiders}: This category is in contrast to \emph{the Object-Orientation Abusers}, since code smells belonging to this category do not (intentionally or unintentionally) apply any object-oriented practice. 
We here included the code smell \textit{Functional Decomposition}.
\end{itemize}

Since three categories (\textit{Change Preventers}, \textit{Encapsulators}, \textit{Object-Orientation Avoiders} are based on only one code smell, we did not analyze them independently since they will provide the same results as those of the code smells belonging to them.
In Table~\ref{tab:addlabel}, we propose a summary of the new  revisited classification of the smells with all the categories we considered and the smells included in each category. In the table, we outline in \emph{italics} the new smells we introduced in the categories of  M\"{a}ntyl\"{a} according to our evaluation and the new category we defined.

{
\centering
\setlength{\tabcolsep}{4pt}
\begin{table}[ht]
  \centering
  \caption{Code Smell Taxonomy}\label{tab:addlabel}%
\footnotesize
\begin{tabular}{ll}
    \hline
    \multicolumn{1}{c}{\textbf{Category Name}} & \multicolumn{1}{c}{\textbf{Code Smells}} \\
    \hline
    \multirow{6}[2]{*}{The Bloaters} & Blob \\
          & Large Class \\
          & Long Method \\
          & Long Parameter List \\
          & \textit{Complex Class} \\
          & \textit{Swiss Army Knife} \\
        \hline
          The Change Preventers & \textit{Spaghetti Code} \\
          \hline
    \multirow{4}[2]{*}{The Dispensables} & Lazy Class\\
          & Speculative Generality\\
          &  \textit{Many Field Attributes But Not Complex}\\
          &  Duplicated Code\\
    \hline
    The Encapsulators & \textit{Message Chain}\\
    \hline
    \multirow{6}[2]{*}{The Object-Orientation Abusers} & Anti-Singleton \\
           & Refused Parent Bequest \\    
          &  \textit{Base Class Knows Derived Class} \\
          &  \textit{Base Class Should Be Abstract} \\
          &  \textit{Class Data Should Be Private} \\
          &  \textit{Tradition Breaker} \\
       \hline
{The Object-Orientation Avoiders} & \textit{Functional Decomposition} \\   
    \hline
    \end{tabular}
\end{table}
}

\pagebreak
\subsection{Architectural Smells}
\label{sec:Background_ASmells}
The architectural smells we considered in our study are those described below, where a subsystem (component) refers to a set of packages and classes identifying an independent unit of the system responsible for a certain functionality:
\begin{enumerate}
  \item  \emph{Unstable Dependency (UD)}: describes a subsystem (component) 
  that depends on other subsystems that are less stable than itself \cite{Martin2007}. 
  This may cause a ripple effect of changes in the   system~\cite{ArcelliFontana2016c}. Detected in packages.
   \item\emph{Hub-Like Dependency (HD)}: arises when an abstraction has (outgoing and incoming) dependencies on a large number of other abstractions~\cite{Suryanarayana2015}. 
   Detected in classes and packages.
  \item  \emph{Cyclic Dependency (CD)}: refers to a subsystem (component) that is involved in a chain of relations that break the desirable acyclic nature of a subsystem's dependency structure. 
  The subsystems involved in a dependency cycle are hard to release, maintain, or reuse in isolation. 
  Detected  in classes and packages. 
  The \emph{Cyclic Dependency} AS is detected according  to  different shapes~\cite{DBLP:conf/aswec/Al-MutawaDMM14} as described in~\cite{ArcelliFontana2017}.

  \item  \emph{Multiple Architectural Smell (MAS)}: identifies a subsystem (component) that is affected by more than one architectural smell and provides the number of the architectural smells involved. 
  
\end{enumerate}
We decided to consider these AS in the study since 
they represent relevant  problems related to dependency issues: Components with high coupling and
a large number of dependencies cost more to maintain and
hence can be considered more critical, leading to a progressive architectural degradation \cite{Ernst2015}. In particular, Cyclic Dependency is one of the most common architectural smells that is dangerous and difficult to remove \cite{Martini2018}. Moreover, a tool called Arcan that can detect these smells is  available. As outlined in Section~\ref{sec:background_as}, few tools for AS detection are currently freely available. Other AS impacting different issues will be considered in the future as their automatic detection will become possible.

\section{Case Study Design}
\label{sec:CaseStudy}
The goal of our work is to understand whether architectural smells could be derived and obtained from code smells or whether they are independent from them. For this purpose, we conducted a case study to investigate the interdependency between architectural smells and code smells by analyzing  111 open-source Java projects. For the design and conduction of the case study, we followed the guidelines proposed by Runeson~\cite{Runeson2009}. 

In this section, we will present the goal, the research questions, the metrics, and the hypotheses for the case study. Based on them, we will outline the study context, the data collection, and the data analysis.

\subsection{Goal, Research Questions, Metrics, and Hypotheses}
We formulated our goal according to the GQM approach~\cite{Basili1994}\\ 
\textit{Analyze} code smells and architectural smells \\
\textit{for the purpose of} evaluating them\\
\textit{with respect to} their interdependency\\
\textit{from the point of view of} developers\\
\textit{in the context of} open-source Java projects.\\

Based on our goal, we derived the following Research Questions (RQ), Metrics (M), and Hypotheses (H)~\cite{Basili1994}, \cite{Shull2007}.

\textbf{RQ1}: Is the presence of an architectural smell independent from the presence of code smells? 
\begin{itemize}
\item M1: correlation coefficient between architectural smells and code smells
\begin{itemize}
\item H0: The presence of an architectural smell is independent from the presence of code smells. 
\item H1: The presence of an architectural smell depends on the presence of code smells. 
\end{itemize}
\end{itemize}

\textbf{RQ1.1}: Is the presence of a Multiple Architectural Smell (MAS) independent from the presence of code smells? 
\begin{itemize}
\item M1.1: correlation coefficient between Multiple Architectural Smell and code smells.
\begin{itemize}
\item H0: The presence of a Multiple Architectural Smell (MAS) is independent from the presence of code smells. 
\item H1: The presence of a Multiple Architectural Smell (MAS) depends on the presence of code smells. 
\end{itemize}
\end{itemize}

\textbf{RQ2}: Is the presence of an architectural smell independent from the presence of a \emph{category} of code smells?  
\begin{itemize}
\item M2: correlation coefficient between architectural smells and categories of code smells.
\begin{itemize}
\item H0: The presence of an architectural smell is independent from the presence of a \emph{category} of code smells.
\item H1: The presence of an architectural smell depends on the presence of a \emph{category} of code smells. 
\end{itemize}
\end{itemize}

\textbf{RQ2.1}: Is the presence of a Multiple Architectural Smell independent (MAS) from the presence of a \emph{category} of code smells? 
\begin{itemize}
\item M2.1: correlation coefficient between Multiple Architectural Smell and categories of code smells.
\begin{itemize}
\item H0: The presence of  a Multiple Architectural Smell (MAS) is independent from the presence of a \emph{category} of code smells.
\item H1: The presence of  a Multiple Architectural Smell (MAS) depends on the presence of a \emph{category} of code smells.
\end{itemize}
\end{itemize}

With our RQs, we aim to understand whether a single architectural smell (\textbf{RQ1}) or a Multiple Architectural Smell (\textbf{RQ1.1}) can be independent  from code smells or from a category that groups code smells as described in Section 3.2 (\textbf{RQ2} and \textbf{RQ2.1}). 

\subsection{Study Context}
\label{context}
We selected projects contained in the Qualitas Corpus collection of software projects ~\cite{Tempero2010}. In particular, we used the compiled version of the Qualitas Corpus ~\cite{Terra2013}. 111 Java projects are available and already compiled with more than 18 million LOCs, 16,000 packages, and 200,000 classes analyzed. The data set includes projects from different contexts such as IDEs, SDKs,  databases, 3D/graphics/media, diagram/visualization libraries and tools, games, middlewares, parsers/generators/make tools, programming language compilers, testing libraries and tools, and other tools not belonging to the previous categories. Terra et al.~\cite{Terra2013} provide more information on the context and types of these projects. 
 


\subsection{Data Collection}
\begin{figure}[h!]
\centering
\includegraphics[trim=11mm 92mm  50mm 20mm,  clip, width=0.9\linewidth]{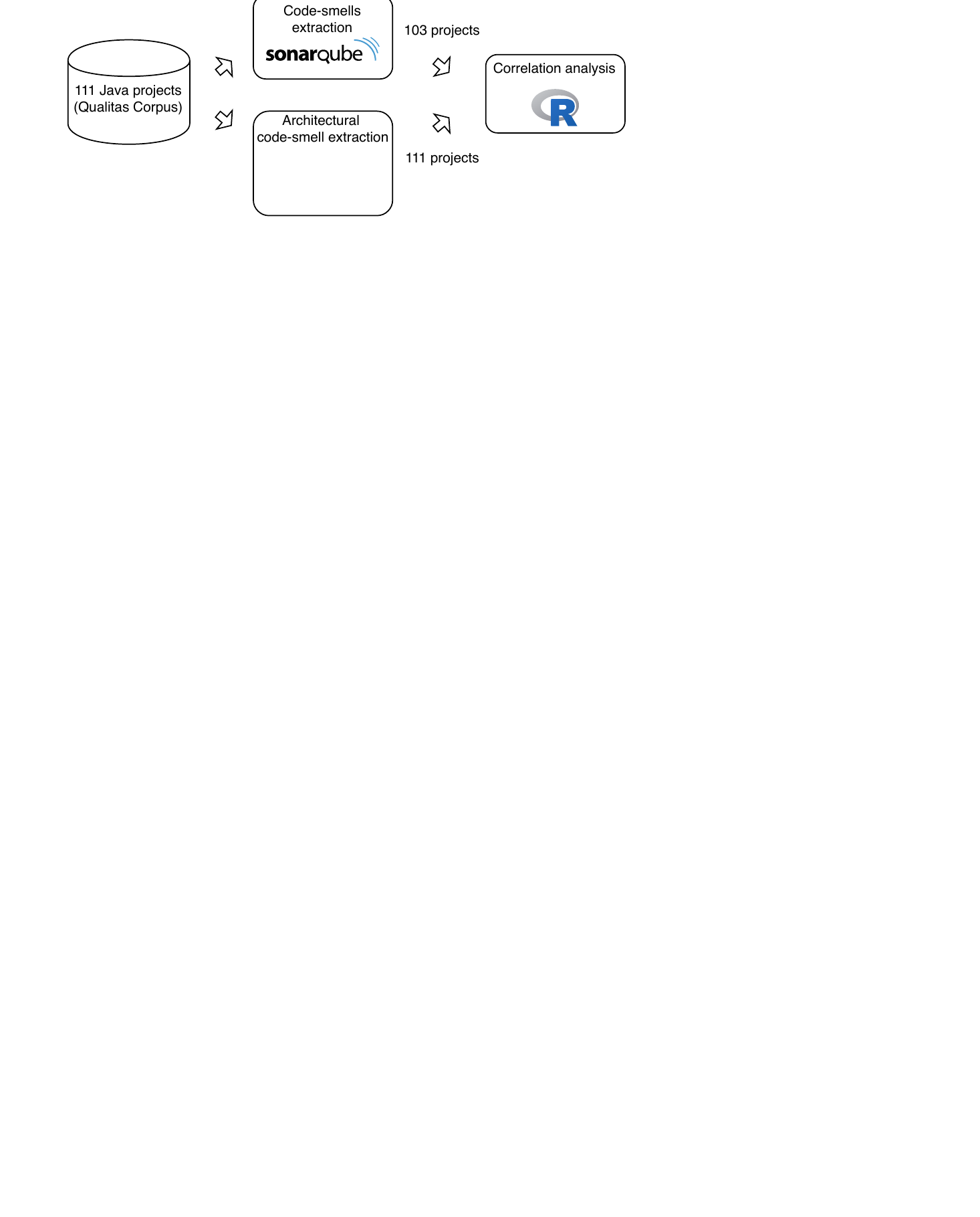}
\caption{Data Collection Process and Data Analysis}
\label{fig:DataCollection}
\end{figure}

We detected architectural smells in 111 Java projects  and code smells in 103 Java projects of the Qualitas Corpus~\cite{Terra2013}, as depicted in Figure~\ref{fig:DataCollection}.
%
%

Architectural smells were detected in these projects through the Arcan tool ~\cite{ArcelliFontana2017}, while the analysis of code smells was carried out with SonarQube  using the ''Antipatterns-CodeSmell'' plugin.
The results of this step are lists of the architectural smells and code smells present in each analyzed project. The raw data is available in the replication package~\cite{TaibiRawData2017}.

\subsubsection{Code smell detection data}
The SonarQube ''Antipatterns-CodeSmell'' plugin is a code smell detection tool that integrates DECOR (Defect dEtection for CORrection) ~\cite{DECOR} into SonarQube, detecting the 19 code smells reported in Section~\ref{sec:Background_smells}. 
DECOR can be applied to any object-oriented language; however, the SonarQube plugin is only configured to detect code smells in Java. Moreover, SonarQube also calculates several other static code metrics such as the number of lines of code and cyclomatic complexity, but also reports code violations. 

It is important to note that in SonarQube (up to the  version 6.5), the term ''Code Smells'' is used to report coding style violations (also known as Issues in SonarQube), such as brackets closed on the wrong line, or redundant throw declarations.
To avoid misunderstandings with coding style violations, the SonarQube ''Antipatterns-CodeSmell'' plugin tags all the code smells of Section~\ref{sec:Background_smells} as ''Antipatterns/CodeSmells''. 
Regarding detection accuracy, we relied on the DECOR  detection tool since it ensures 100\% recall for the detection of code smells~\cite{DECOR}. 
Moreover, since the  definition of code smells is based on several metrics and thresholds, we relied on the standard metrics  proposed by Moha et al.~\cite{DECOR} so as to ensure a  precision average of 80\%.

The detection of code smells in the Qualitas Corpus data set was carried out on a Linux virtual machine with 4 cores and 16GB of RAM. The first 103 projects were analyzed within 35 days. Due to time constraints, we skipped the analysis of the remaining eight projects such as Eclipse and JBoss, which would have taken more than three months. The reason for this dramatic increase in analysis time is due to the project structure. These eight projects are composed of several sub-projects with sizes similar to the other 103 projects already analyzed. Therefore, in this work we only consider the results of the 103 projects listed in \ref{sec:appendixa}.

\subsubsection{Architectural smell detection data}
The Arcan tool  focuses on the identification of architectural smells whose generation was caused by instability issues. By software instability we mean the inability to make changes without impacting  the entire project or a large part of it. 
To accomplish its aim, the tool computes the metrics proposed by Martin~\cite{Martin1995} and exploits them during the analysis.
The detection techniques exploit graph databases to perform graph queries, which allows higher scalability in the detection and management of a large number of different kinds of dependencies.

The detection techniques for AS and the validation of the tool results have been described in previous studies 
~\cite{ArcelliFontana2016c}, \cite{ArcelliFontana2017}.  The results of the tool were validated on ten open-source projects and two industrial projects based on feedback from the developers with a high precision value of 100\% and  a recall value of 66\%. The developers also reported five  architectural smells that were false negatives, but these cases were related to external components beyond the scope of the analysis performed by the tool.
Moreover, the results of Arcan were evaluated using the feedback of practitioners in four industrial projects \cite{Martini2018}.


In this study, the detection of the architectural smells was performed on a Windows machine with 4 cores and 24 GB of RAM. The entire Qualitas Corpus data set was analyzed using Arcan within less than 24 hours. 
The tool is freely available and easy to install and use  \footnote{http://essere.disco.unimib.it/wiki/arcan)}.

\subsection{Data Analysis}
\label{sec:DataAnalysis}
In this section, we will describe the procedure we followed to analyze the collected data in order to answer our research questions.

We analyzed the classes infected both by an architectural smell and  one or more code smells at the class and package levels.

Architectural smells involve more than one Java class, while the 19 code smells considered in this work involve only one class. Therefore, for each architectural smell, we could have one or more code smells infecting the same set of classes. 
In the analysis, we only calculated correlations between code smells infecting those classes (and packages) that were also infected  by architectural smells. 

To give an example: \emph{Classes A}, \emph{B}, and \emph{C} may be infected by Cyclic Dependency, while \emph{classes A} and \emph{C} may be infected by  God Class and \emph{class D} may be infected by Speculative Generality. 
In this case, we would calculate the correlation only for the architectural smell Cyclic Dependency and the code smell God Class, since they affect the same set of classes, whereas we would not consider the code smell Speculative Generality, since it infects a class that is not infected by Cyclic Dependency. 

Before answering our RQs, we analyzed the distribution of the code smells and the architectural smells in our data set. We performed a descriptive analysis of the collected data, analyzing the number of code smells and architectural smells per project and per package.

We analyzed the frequency of occurrence of the code smells and architectural smells,  considering:
\begin{itemize}
\item \textbf{(CS+AS)}: Classes infected by code smells \textbf{AND} architectural smells;

\item \textbf{(CS)}: Classes infected \textbf{only by} code smells;

\item \textbf{(AS)}: Classes infected \textbf{only by} architectural smells;

\item \textbf{(HC)}: Healthy Classes -- classes neither infected by code smells nor by architectural smells.
\end{itemize}

We analyzed the 103 projects independently, then considered the data of all the projects globally, as though all the classes belonged to one single project.
Projects without code smells or architectural smells were not considered for the analysis. 

\begin{figure}[H]
\centering
\includegraphics[trim={3cm 21.5cm 3cm 2cm},clip,width=\linewidth]{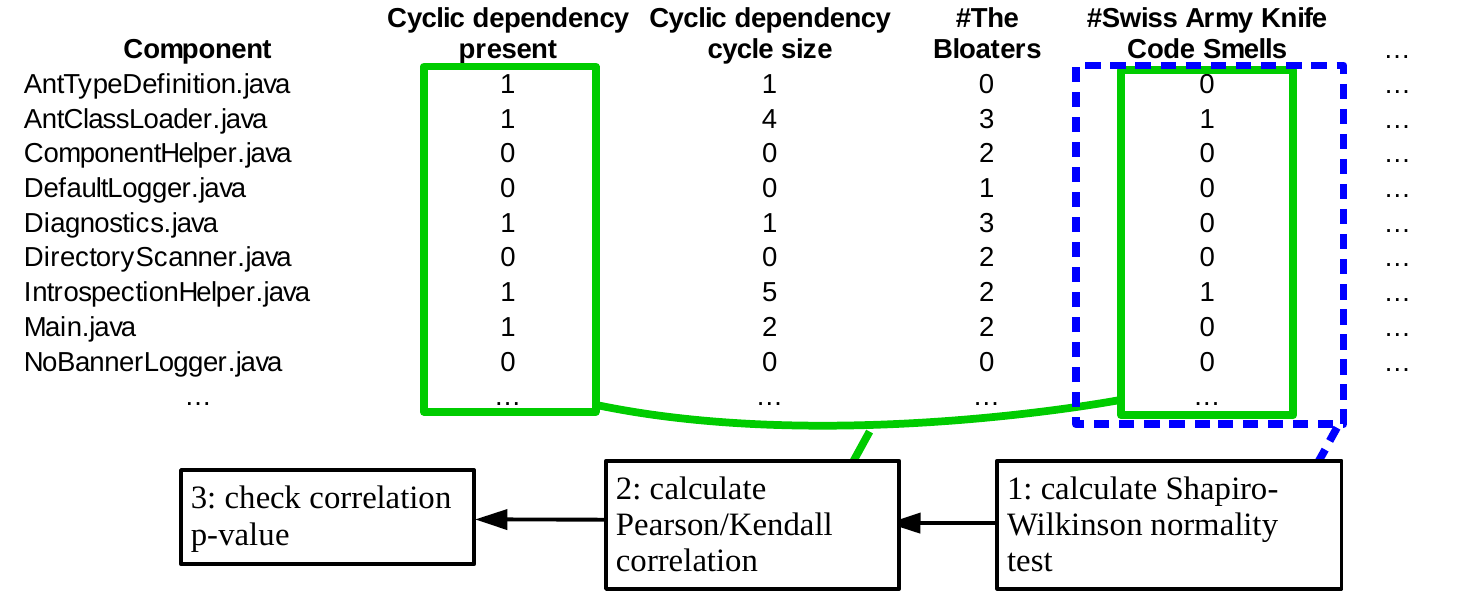}
\caption{The Data Analysis Process}
\label{fig:DataAnalysis}
\end{figure}

In order to answer  our research questions, we applied the following analysis procedure, as summarized in Figure~\ref{fig:DataAnalysis}.
We considered as our \textit{dependent variable} the number of each type of  architectural smell infecting the same classes and as \textit{independent variable} the number of code smells infecting the same classes. 
We investigated the correlation for every pair of (code smell and architectural smell or categories of code smells and architectural smell), since considering all types of smells at the same time might hide possible correlations among smells, making it impossible to discover them.

\begin{itemize}
\item For each Architectural Smell
\begin{itemize}
\item \textit{Data-Normality Test}: We tested the  data for normality by means of the Shapiro-Wilk test.
\item \textit{Correlation Analysis}: We calculated the correlation between code smells or a category of smells (independent variable) and architectural smells or Multiple Architectural Smells (dependent variable). 
\begin{itemize}
\item If the data were normally distributed, we calculated the Pearson correlation coefficient
\item If the data were not normally distributed,  we calculated the Kendall rank correlation coefficient.
\end{itemize}
\end{itemize}
\end{itemize}

Correlation is a bi-variate analysis that measures the association strength between two variables and the direction of the relationship. The value of the correlation coefficient varies between +1 and -1, where a value of ± 1 means a perfect degree of association between the two variables.  

Usually, in statistics, different types of correlations are applied. Pearson correlation is the one used most frequently to measure the relationship degree between linearly related variables. 
Kendall rank correlation is one of the non-parametric tests commonly used to measure the strength of dependency between two variables. We selected Kendall rank correlation because compared with other non-parametric tests, it has less gross error sensitivity (GES), meaning more robustness, and a smaller asymptotic variance (AV), meaning more efficiency~\cite{Croux2010}.

We only show those results with a p-value smaller than 0.05 as a statistical significance threshold. This is customary in Empirical Software Engineering studies~\cite{Shull2007}.


\section{Results}
\label{sec:Results}
In this section, we will first describe the data we analyzed and then  answer  our research questions by reporting the results of the analysis described in Section~\ref{sec:DataAnalysis}.

All the projects contain classes infected by both architectural smells and code smells.

Considering the presence of code smells in the 103 projects, only 15 of the 19 code smells detectable by the SonarQube plugin were found. The 103 projects were not infected by Blob Class, Functional Decomposition, Base Class Knows Derived, and Tradition Breaker. 
This also impacted  the categories of code smells containing code smells not found in the  projects, since two categories  (Change Preventers and Object Orientation Avoiders) were based on two code smells not detected in the 111 projects. Therefore, only the remaining four categories of code smells are considered in the analysis. 

Regarding the architectural smells, Arcan detected them in 102 projects (Jasml contains no architectural smell). 
Therefore, we considered this set of 102 projects for the analysis. 
Note that, for the sake of completeness, we also report data for the code smell categories containing only one code smell. 
However, these categories will not be considered in the next analysis to avoid duplication of the results. 

Table~\ref{tab:SmellsPerProject} shows the number of projects infected by code smells, categories of code smells, and architectural smells (Column \#Inf.prj.), while the remaining columns report descriptive statistics. 
Regarding code smells, Complex Class, Long Method, and Long Parameter List were the most commonly detected ones in the projects (more than 100 projects). 
Swiss Army Knife, Message Chains, and Large Class were code smells infecting fewer projects (less than 11), while Base class Knows Derived Class, Blob Class, Functional Decomposition, and Tradition Breaker were not present in any of the analyzed projects. 

\begin{figure}[H]
\centering
\includegraphics[width=\linewidth]{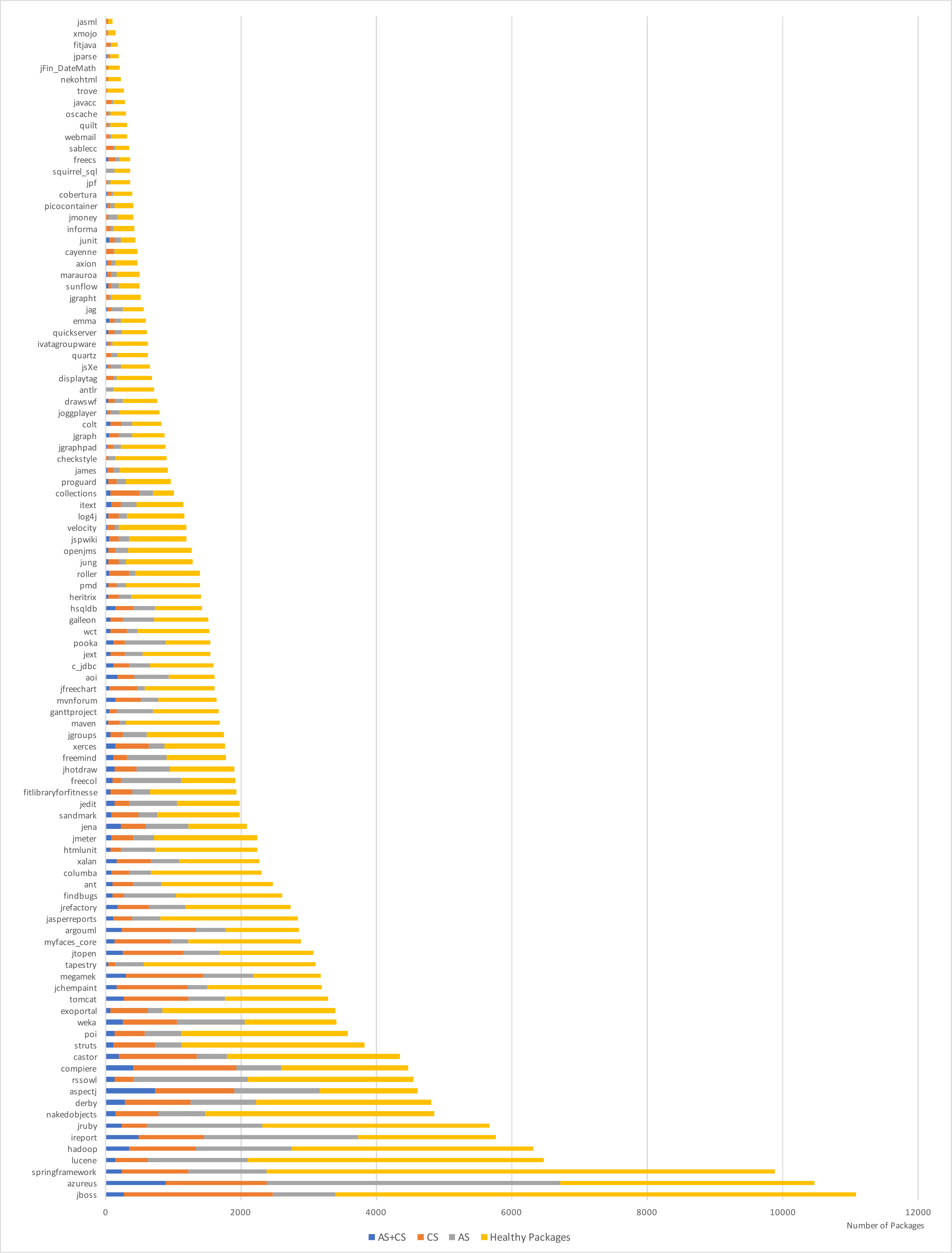}
\caption{Number of packages infected by code smells or architectural smells}
\label{fig:AsCsDistribution}
\end{figure}

\begin{figure}[H]
\centering
\includegraphics[width=\linewidth]{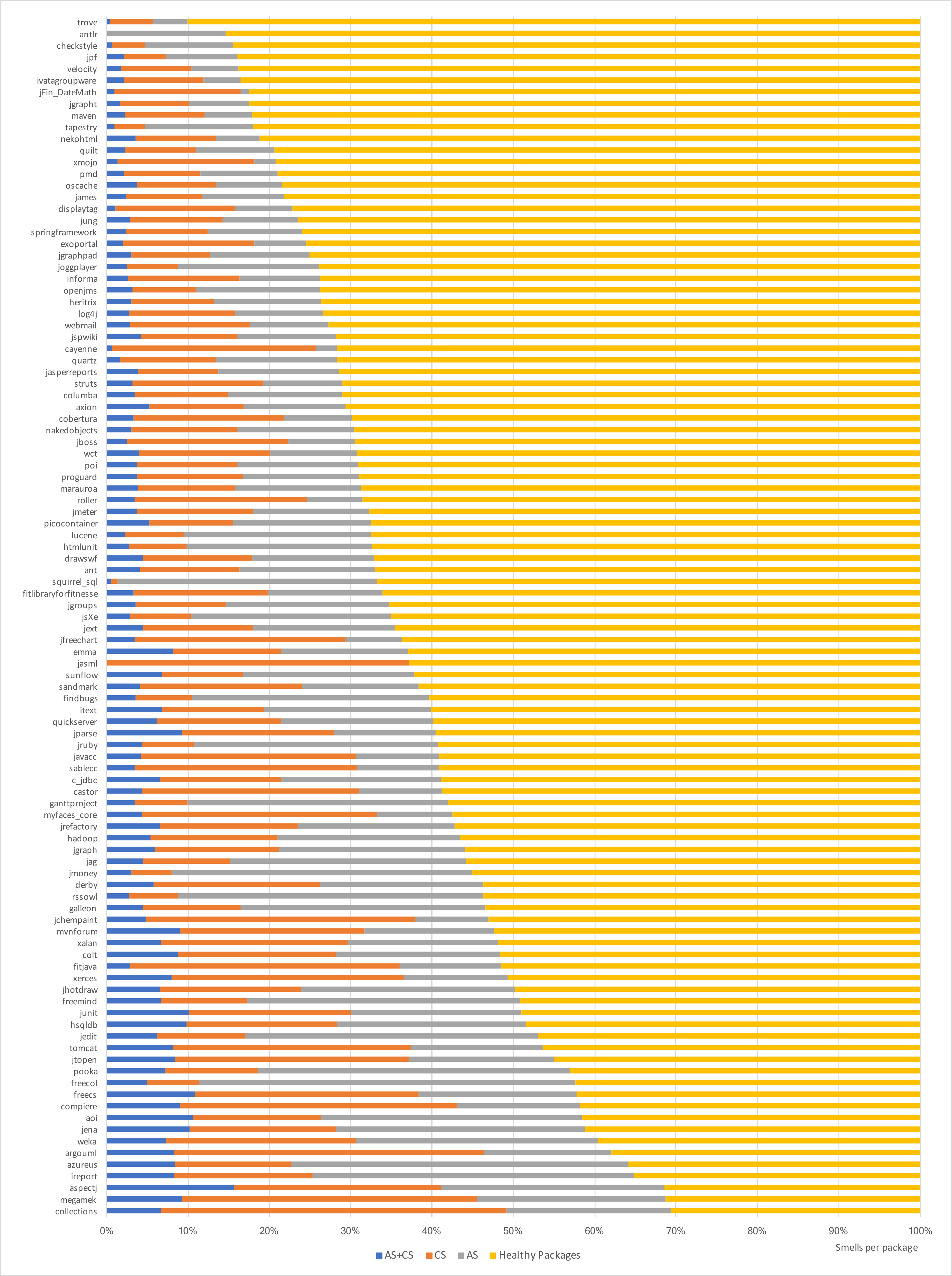}
\caption{Number of code smells and architectural smells per package}
\label{fig:SmellsPerPAckage}
\end{figure}

Figure~\ref{fig:AsCsDistribution}
shows the number of classes infected by code smells and architectural smells (CS+AS), classes infected only by code smells (CS), classes infected only by architectural smells (AS), and healthy classes, i.e., classes without any smells (HC) in our data set. Moreover, Figure~\ref{fig:SmellsPerPAckage} shows the distribution of the same data per package.


Regarding the architectural smells, all the projects were infected by at least two architectural smells. The analysis revealed that 101 projects were infected by Cyclic Dependency, 100 were infected by Hub-Like Dependency, 95 were infected by Unstable Dependency, and 102 were infected by a Multiple Architectural Smell.

Table~\ref{tab:ProjectsInfectedDetails} (\ref{sec:appendixa}) reports the details on the number of code smells and architectural smells detected in each project. 

In Table~\ref{tab:ProjectsInfectedbyCD}, Table~\ref{tab:ProjectsInfectedbyHD}, Table~\ref{tab:ProjectsInfectedbyUD}, and Table~\ref{tab:ProjectsInfectedbyMAS}, we report the results obtained from analyzing the AS-CS pairs, while in Table~\ref{tab:ProjectsInfected2} we present the results for the  AS-CS category pairs. These tables report the number of infected projects for each pair (column ``\textit{\#Inf. Prj.}''), the number of infected projects where the results are statistically significant and their percentage up to the total number of infected projects (column ``\textit{\#Prj.(p$<$0.05)}''). Moreover, we also list the projects that reported a Kendall correlation  higher than 0.5 (column ``\textit{\#Prj.(tau$<$0.5)}''). 

As an example (Table~\ref{tab:ProjectsInfectedbyUD}), the pair composed of the architectural smell \textit{Unstable Dependency} (UD) and the code smell \textit{Base Class Should be Abstract} (BCSA)  was detected in 54 projects  (column ''\textit{\#Inf.prj}''), with 30 of them (55\% of projects) having a significant statistical correlation with a p-value $<$0.05 (column \textit{''\#Prj.(p-value$<$0.05''}). However,  only two projects have a correlation higher than 0.5 (column ``\textit{\#Prj. (tau $>$0.5)}'') while the remaining ones (28 projects), which are not listed in the table,  had a statistically significant result with a low correlation (tau $<$0.5). The column ``\textit{Project}'' indicates the two projects with a correlation higher than 0.5. 

We also performed the same analysis (AS-CS pairs and AS-CS category pairs) at the project level, trying to analyze all the classes together as  belonged to a single project. The results did not change, as illustrated in Table~\ref{tab:GloballyRQ1} and Table~\ref{tab:GloballyRQ2}.
We report the correlation value (column ''\textit{tau}'') and the relative  statistical hypothesis testing value (column ''\textit{p-value}'').

{
\centering
\setlength{\tabcolsep}{4pt}
\begin{table}[!h]
\captionsetup[table]{skip=5pt}
\caption{Projects infected by code smells, a category of code smells, or architectural smells} 
\label{tab:SmellsPerProject} 
\footnotesize
\begin{tabular} {@{}p{5cm}|p{1cm}|p{1.2cm}p{1.1cm}p{0.8cm}p{1.4cm}@{}}
\hline
 \multirow{2}{*}{\textbf{Name}} & \textbf{\#Inf.}  & \multicolumn{4}{c}{\textbf{per Project}}\\  \cline{3-6}
& \textbf{prj.}& \textbf{AVG} & \textbf{Max} & \textbf{Min} & \textbf{StD}\\
\hline
\textbf{Code Smells} & \\
\hspace{4mm}Complex Class & 103 & 147.90 & 914 & 1 & 163.23 \\
\hspace{4mm}Duplicated Code & 103 & 237.28 & 1830 & 0 & 357.67\\
\hspace{4mm}Long Method & 102 & 178.88 & 1,251 & 0 & 197.58\\
\hspace {4mm}Long Parameter List & 100 & 94.09 & 1,197 & 0 & 157.51\\
\hspace {4mm}Anti-Singleton & 92 & 31.96 & 7.34 & 0 & 81.86\\
\hspace {4mm}Class Data should be Private & 90 & 28.93 & 3.53 & 0 & 50.07\\
\hspace{4mm}Lazy Class & 86 & 26.96 & 210 & 0 & 43.65\\
\hspace {4mm}Spaghetti Code & 58 & 2.97 & 40 & 0 & 5.23 \\
\hspace {4mm}Baseclass Abstract & 54 & 3.84 & 65 & 0 & 8.49\\
\hspace {4mm}Refused Parent Bequest & 42 & 6.38 & 139 & 0 & 19.33\\
\hspace {4mm}Speculative Generality & 36 & 2.68 & 35 & 0 & 5\\
\hspace {4mm}Many Field Attr. not Complex & 32 & 0.76 & 20 & 0 & 2.23\\
 \hspace {4mm}Swiss Army Knife & 11  & 1.39 & 76 & 0 & 8.22\\
\hspace {4mm}Message Chains & 8 & 1.27 & 62 & 0 & 7.19\\
\hspace{4mm}Large Class & 5 & 0.07 & 2 & 0 & 0.32\\
\hspace {4mm}Baseclass Knows Derived & 0 & - & - & - & -\\
\hspace {4mm}Blob Class & 0 & - & - & - & -\\
\hspace {4mm}Functional Decomposition & 0 & - & - & - & -\\
\hspace {4mm}Tradition Breaker & 0 & - & - & - & -\\
\hline
\textbf{Category of Code Smells} & \\
The Bloaters & 103 & 421.70 & 3,364 & 1 & 496.13\\
\hspace {4mm}The Dispensables & 102 & 264.24 & 1,849 & 0 & 379.22\\
\hspace {4mm}The Obj.-Orientation Abusers & 92 & 31.96 & 734 & 0 & 81.86 \\
\hspace {4mm}The Change Preventers & 58 & 2.97 & 40 & 0 & 5.23 \\
\hspace {4mm}The Encapsulators	& 8 & 1.27 & 62 & 0 & 7.19\\
\hspace {4mm}The Obj.-Orientation Avoiders	 & 0 & - & - & - & -\\
\hline
\textbf{Architectural Smells} &\\
\hspace {4mm}Multiple Architectural Smell & 102 & 6,148.02 & 162,531 & 0 & 22,176.7\\
\hspace {4mm}Cyclic Dependency & 101 & 6,122.24 & 162,357 & 0 & 22,162.1\\
\hspace {4mm}Hub-Like Dependency & 100 & 21.35 & 168 & 0 & 25.43\\
\hspace {4mm}Unstable Dependency & 95 & 4.43 & 15 & 0 & 3.16\\
\hline
\end{tabular}
\end{table}
}

In Table~\ref{tab:ProjectsInfectedDetails} (\ref{sec:appendixa}), we report the number of architectural smells, categories of code smells, and code smells infecting each analyzed project.

In order to better understand the cases of positive correlations, we manually inspected all the 23 projects where we found pairs with a correlation higher than 0.5 with a p-value lower than 0.05. The result of the manual inspection did not yield any useful feedback. As an example, Anti-Singleton (ASG) is positively correlated with Cyclic Dependency only in the project xmojo. Manually inspecting its classes, we confirmed the presence of the four cyclic dependencies, where two cycles included one class per cycle also affected by ASG and one of the four cycles was also affected by a Spaghetti Code smell. The same class affected by Spaghetti Code was also affected by Hub-Like Dependency (HD). 
Other projects, such as Checkstyle, JParse, and Log4J reported a relatively higher number of AS and CS but their manual examination did not reveal any noticeable information.

{
\centering
\setlength{\tabcolsep}{4pt}
\begin{table}[ht]
\captionsetup[table]{skip=5pt}
\caption{Projects infected by the Cyclic Dependency architectural smell (CD) and code smells (RQ1)} 
\label{tab:ProjectsInfectedbyCD} 
\footnotesize 
\begin{tabular}[h]
{@{}p{1cm}|p{1.2cm}|p{1.8cm}|p{1.7cm}|p{0.5cm}|p{1cm}|p{3.5cm}@{}}
\hline
\multirow{2}{*}{\textbf{AS}} & \multirow{2}{*}{\textbf{CS}} & \multirow{2}{*}{\textbf{\#Inf.prj}} &  
\multicolumn{2}{|c|}{\textbf{Prj.(p-value$<$0.05)}} & \multicolumn{2}{|c}{\textbf{Prj.(tau$>$0.5)}} \\ \cline{4-7}
& & &\# &\% &\# & prj. name  \\ \cline{1-7}
\multirow{15}{*}{CD} & ASG & 92 & 70 & 76 & 1 & xmojo \\
& BCSA &  54 &  45 &  83 & 0 & - \\
& CC & 102 & 92 & 90 & 1 & freecs \\
& DC & 102 &  87 & 85 & 0 & - \\
& DsP &  90 & 60 &  67 & 0 & - \\
& LC &  5 & 1 & 20 &  0 & - \\
& LM &  102 &  87 & 85 & 0 & - \\
& LPL & 100 & 80 & 80&  1 & jparse  \\
& LzC & 86 &  28 &32   &  0 & -  \\
& MFnC &  32 &  10 & 31  & 0 & - \\
& MC &  8 &  7 & 87  &  0 & - \\
& RPB &  42 &  26 & 62  & 0 & - \\
& SC & 58 & 40 & 69 & 1 & xmojo \\
& SG &  36 &  22 & 61  & 0 & - \\
& SAK &  11 &  6 &  54 & 0 & - \\
\hline 
\end{tabular}
\end{table}
}

{
\centering
\setlength{\tabcolsep}{4pt}
\begin{table}[H]
\captionsetup[table]{skip=5pt}
\caption{Projects infected by the Hub-like Dependency architectural smell (HD) and code smells (RQ1)} 
\label{tab:ProjectsInfectedbyHD} 
\footnotesize 
\begin{tabular}[h]
{@{}p{1cm}|p{1.2cm}|p{1.8cm}|p{1.7cm}|p{0.5cm}|p{1cm}|p{3.5cm}@{}}
\hline
\multirow{2}{*}{\textbf{AS}} & \multirow{2}{*}{\textbf{CS}} & \multirow{2}{*}{\textbf{\#Inf.prj}} &  
\multicolumn{2}{|c|}{\textbf{Prj.(p-value$<$0.05)}} & \multicolumn{2}{|c}{\textbf{Prj.(tau$>$0.5)}} \\ \cline{4-7}
& & &\# &\% &\# & prj. name  \\ \cline{1-7}
\multirow{15}{*}{HD} &ASG & 92 & 80 & 89 & 1 & jmoney \\ 
& BCSA & 54 & 50 & 92 & 2 & checkstyle, jparse \\
& CC & 102 & 95 & 90 & 0 & - \\
& DC & 102 & 91 & 89 & 0 & - \\
& DSP & 90 & 80 & 89 & 2 & checkstyle, jparse \\
& LC &  5 &  5 &  100 &  0 & - \\
& LM &  102 &  94 & 94 &0  & - \\
& LPL & 100 & 93 & 93 & 0 & - \\
& LzC &  86 &  78 & 91 & 0 & - \\
& MfNC & 32 & 26 & 81 & 1 & checkstyle \\
& MC &  8 &  7 & 87 & 0 & - \\
& RBP & 42 & 37 & 88 & 1 & checkstyle \\
& SC & 58 & 50 & 69 & 1 & xmojo \\
& SG &  36 &  31 &  86 & 0 & - \\
& SAK &  11 &  9 &  82 & 0 & - \\
\hline 
\end{tabular}
\end{table}
}

{
\centering
\setlength{\tabcolsep}{4pt}
\begin{table}[H]
\captionsetup[table]{skip=5pt}
\caption{Projects infected by the Unstable Dependency architectural smell (UD) and code smells (RQ1)} 
\label{tab:ProjectsInfectedbyUD} 
\footnotesize 
\begin{tabular}[h]
{@{}p{1cm}|p{1.2cm}|p{1.8cm}|p{1.7cm}|p{0.5cm}|p{1cm}|p{3.5cm}@{}}
\hline
\multirow{2}{*}{\textbf{AS}} & \multirow{2}{*}{\textbf{CS}} & \multirow{2}{*}{\textbf{\#Inf.prj}} &  
\multicolumn{2}{|c|}{\textbf{Prj.(p-value$<$0.05)}} & \multicolumn{2}{|c}{\textbf{Prj.(tau$>$0.5)}} \\ \cline{4-7}
& & &\# &\% &\# & prj. name  \\ \cline{1-7}
\multirow{15}{*}{UD} & ASG & 92 & 60 & 65 & 1 & nekohtml\\
& BCSA & 54 & 30 & 55 & 2 & log4j, picocontainer \\
& CC & 102 & 92 & 90 & 0 & - \\
& DC & 102 & 84 & 82 & 0 & - \\
& DsP & 90 & 63 & 70  & 0 & - \\
& LC &  5 &  4 &  80 & 0 & - \\
& LM &  102 &  82 & 80 & 0 & - \\
& LPL & 100 & 68 & 68 & 0 & - \\
& LzC &  86 &  36 & 42 & 0 & - \\
& MFnC &  32 &  9 & 28 & 0 & - \\
& MC &  8 &  5 & 62 & 0 & - \\
& RBP & 42  &  23 & 55 & 0 & -  \\
& SC & 58 & 30 & 52& 1 & oscache \\
& SG & 36 & 17 &  & 2 & log4j, picocontainer \\
& SAK &  11 &  8 &  72 & 0 & - \\
\hline 
\end{tabular}
\end{table}
}

{
\centering
\setlength{\tabcolsep}{4pt}
\begin{table}[H]
\captionsetup[table]{skip=5pt}
\caption{Projects infected by a Multiple Architectural Smell (MAS) and code smells (RQ1.1)} 
\label{tab:ProjectsInfectedbyMAS} 
\footnotesize 
\begin{tabular}[h]
{@{}p{1cm}|p{1.2cm}|p{1.8cm}|p{1.7cm}|p{0.5cm}|p{1cm}|p{3.5cm}@{}}
\hline
\multirow{2}{*}{\textbf{AS}} & \multirow{2}{*}{\textbf{CS}} & \multirow{2}{*}{\textbf{\#Inf.prj}} &  
\multicolumn{2}{|c|}{\textbf{Prj.(p-value$<$0.05)}} & \multicolumn{2}{|c}{\textbf{Prj.(tau$>$0.5)}} \\ \cline{4-7}
& & &\# &\% &\# & prj. name  \\ \cline{1-7}
\multirow{15}{*}{MAS} & ASG & 92 & 66 & 72 & 0 & - \\
& BCSA & 54 & 32  & 60 & 0 &  -\\
& CC & 102 &  90 & 88 & 0 & - \\
& DC & 102 & 64 & 63 & 0 &  -\\
& DsP & 90 & 56 & 62 & 0 & - \\
& LC & 5 & 0 &  0 & 0 & - \\
& LM & 102 & 83 & 81 &0  & - \\
& LPL & 100  & 80  & 80 & 1 & jparse \\
& LzC & 86 & 33 & 38 & 0 & - \\
& MFnC & 32 & 7 & 22 & 0 & - \\
& MC & 8 & 7 & 87 & 0 &  \\
& RBP & 42 & 21  & 50 & 0 & - \\
& SC & 58 & 36 & 62 & 1 & xmojo \\
& SG & 36 & 21 & 58 &  0 & - \\
& SAK & 11 & 7 & 63 & 0 & - \\
\hline 
\end{tabular}
\end{table}
}

{
\centering
\setlength{\tabcolsep}{4pt}
\begin{table}[H]
\captionsetup[table]{skip=5pt}
\caption{Projects infected by architectural smells (RQ2) or Multiple Architectural Smells (RQ2.1) and by categories of code smells} 
\label{tab:ProjectsInfected2} 
\footnotesize 
\begin{tabular}[h]
{@{}p{1cm}|p{1.3cm}|p{1.8cm}|p{1.7cm}|p{0.5cm}|p{1cm}|p{3.5cm}@{}}
\hline
\multirow{2}{*}{\textbf{AS}} & \multirow{2}{*}{\textbf{CS cat.}} & \multirow{2}{*}{\textbf{\#Inf.prj}} &  
\multicolumn{2}{|c|}{\textbf{Prj.(p-value$<$0.05)}} & \multicolumn{2}{|c}{\textbf{Prj.(tau$>$0.5)}} \\ \cline{4-7}
& & &\# &\% &\# & prj. name  \\ \cline{1-7}
\multirow{3}{*}{CD} & Bloat. & 103 & 91 & 88 & 0 & - \\
& Disp. & 102 & 74 & 72 & 0 & - \\
& OOA & 98 & 73 & 75 & 0 & - \\
\hline
\multirow{3}{*}{HD} & Bloat.& 103 & 95 & 92 & 0 & - \\
& Disp. & 102 & 90 & 88 & 0 & - \\
& OOA & 92 & 87 & 94 & 1 & jmoney\\
\hline
\multirow{3}{*}{UD} & Bloat.& 102 & 87 & 85 & 0 & -\\
& Disp.  & 102 & 68 & 67 & 0 & -  \\ 
& OAA	&	98 & 69 & 70 & 1 & nekohtml \\
\hline
\multirow{3}{*}{MAS} & Bloat. & 102 & 88 & 86 & 1 & jparse\\
& Disp. & 102 & 68 & 67 & 0 & -  \\ 
& OOA & 98 & 66 & 67 & 0 & - \\
\hline 
\end{tabular}
\end{table}
}

\newpage

\newpage
\newpage
{
\centering
\setlength{\tabcolsep}{4pt}
\begin{table}[H]
\captionsetup[table]{skip=5pt}
\caption{Correlation between AS and CS. All projects merged as a single project (RQ1 and RQ1.1)} 
\label{tab:GloballyRQ1} 
\footnotesize 
\begin{tabular}[h]
{@{}p{1cm}|p{1.3cm}|p{1cm}|p{1.3cm}|p{1cm}|p{1.3cm}|p{1cm}|p{1.3cm}|p{1cm}@{}}
\hline 
\multirow{2}{*}{\textbf{CS}} & \multicolumn{2}{c|}{\textbf{CD}} & \multicolumn{2}{c|}{\textbf{HD}} & \multicolumn{2}{c|}{\textbf{UD}} & \multicolumn{2}{c}{\textbf{MAS}} \\ \cline{2-9}
& \textbf{p-value} & \textbf{tau }& \textbf{p-value} & \textbf{tau} & \textbf{p-value} & \textbf{tau} & \textbf{p-value }& \textbf{tau} \\ \hline
ASG & 0.03 & 0.04 & 0.03& 0.32 & 0.04 & 0.14 & 0.00 & 0.19\\ \hline
BCSA & 0.02 & 0.08 &0.02 & 0.20 & 0.01 & 0.13 & 0.12 & 0.09\\ \hline
CC & 0.01 & 0.13  & 0.04 & 0.31 & 0.01 & 0.03 & 0.00 & 0.23\\ \hline
DC & 0.00 & 0.02  & 0.03& 0.27 & 0.00 & 0.26 & 0.00 & 0.16\\ \hline
DsP & 0.03 & 0.06 &0.04 & 0.33 & 0.03 & 0.28 & 0.00 & 0.15\\ \hline
LC  & 0.72 & 0.35 &  0.03& 0.36 & 0.00 & 0.08 & 0.23 & 0.04\\ \hline
LM & 0.01 & 0.14 & 0.03& 0.27 & 0.00 & 0.23 & 0.92 & 0.25\\ \hline
LPL & 0.03 & 0.19 & 0.03& 0.21 & 0.09 & 0.00 & 0.05 & 0.09\\ \hline
LzC & 0.07 & 0.32 &0.04 & 0.27 & 0.08 & 0.42 & 0.00 & 0.15\\ \hline
MFnC & 0.60 & 0.23 &0.04& 0.27 & 0.08 & -0.05 & 0.00 & 0.14\\ \hline
MC & 0.01 & 0.10&0.04& 0.19 & 0.00 & 0.12 & 0.00 & 0.33\\ \hline
RPB & 0.05 & 0.06 &0.04& 0.31 & 0.08 & -0.30 & 0.91 & 0.01\\ \hline
SC & 0.05 & 0.18 &0.03& 0.36 & 0.06 & 0.18 & 0.11 & 0.04\\ \hline
SG & 0.25 & 0.22 &0.04& 0.19 & 0.09 & 0.04 & 0.01 & 0.14\\ \hline
SAK & 0.06 & 0.15&0.02& 0.20 & 0.04 & 0.29 & 1.11 & 0.13\\ \hline
\end{tabular}
\end{table}
}

{
\centering
\setlength{\tabcolsep}{4pt}
\begin{table}[H]
\captionsetup[table]{skip=5pt}
\caption{Correlation between AS and categories of CS. All projects merged as a single project (RQ2 and RQ2.1)} 
\label{tab:GloballyRQ2} 
\footnotesize 
\begin{tabular}[h]
{@{}p{0.7cm}|p{1.2cm}|p{0.6cm}|p{1.2cm}|p{0.6cm}|p{1.2cm}|p{0.6cm}|p{1.2cm}|p{0.6cm}|p{1.2cm}|p{0.5cm}@{}}
\hline 
\multirow{2}{*}{\textbf{AS}} & \multicolumn{2}{c|}{\textbf{Bloat.}} & \multicolumn{2}{c|}{\textbf{Disp.}} & \multicolumn{2}{c|}{\textbf{OAA}} & \multicolumn{2}{c|}{\textbf{Encap.} }& \multicolumn{2}{c}{\textbf{Change prev.}} \\ \cline{2-11}
& \textbf{p-value} & \textbf{tau }& \textbf{p-value} & \textbf{tau }& \textbf{p-value }& \textbf{tau} & \textbf{p-value} & \textbf{tau} & \textbf{p-value} & \textbf{tau}\\ \hline
CD & 0.02 & 0.32 & 0.03 & 0.17 & 0.03 & 0.16 & 0.64 & 0.10 & 0.17 & 0.11\\ \hline
HD  & 0.00 & 0.24 & 0.00 & 0.28 & 0.08 & 0.03 & 0.07 & 0.36 & 0.75 & 0.29 \\ \hline
UD & 0.03 & 0.12 & 0.03 & 0.14 & 0.05 & 0.19 & 0.21 & 0.05 & 0.87 & 0.26\\ \hline
MAS & 0.06 & 0.23 & 0.11 & 0.13 & 0.16 & 0.17 & 0.33 & 0.00 & 0.28 & 0.05\\ \hline
\end{tabular}
\end{table}
}

\section {Discussion}
\label{sec:Discussion}
In this Section, we will answer our Research Questions (RQs) based on the results obtained and described in Section~\ref{sec:Results} and derive the main lessons learned of this work. 

\subsection{RQ1:Is the presence of an architectural smell independent from the presence of  code smells?}
The results for RQ1 are presented in Table~\ref{tab:ProjectsInfectedbyCD}, Table~\ref{tab:ProjectsInfectedbyHD}, Table~\ref{tab:ProjectsInfectedbyUD}, and Table~\ref{tab:GloballyRQ1}. We analyzed 45 combinations (AS-CS pairs composed of three architectural smells and 15 code smells ) for each of the 102 projects for a total of 4,590 analyses and for the data of all the projects merged together as a single project.  

Regarding the analysis performed separately on 111 projects, we decided not to consider the combinations (CD-LC, CD-MC, CD-SAK), (HD-LC, HD-MC and HD-SAK), and (UD-LC, UD-MC and UD-SAK), due to the low number of infected projects (less than twelve).  
We found statistically significant results (p-value $<$0.05) for all the other combinations. 

However, only 14 combinations in nine projects showed a correlation higher than 0.5. Moreover, the same combination of code smells and architectural smells was found in a maximum of two projects. 
For the other 40 combinations, we found low correlations (tau$<$0.5) in a considerable number of projects (at least 32). 

According to our results, the most interesting low correlations  we found are related to SC (Spaghetti Code)  CS with all the four AS considered, and  ASG (Anti-Singleton) CS  with three AS (CD, HD and UD). 

Considering all the projects merged together, the results did not change and we found for the majority of the cases that the results were not statistically significant.

The results confirm our Hypothesis 0 since, based on the results of the 102 analyzed projects and the analysis of all the projects merged together as a single project, we were unable to identify any dependencies between architectural smells and code smells.

\subsection{RQ1.1:Is the presence of a Multiple Architectural Smell independent from the presence of code smells?}
The results for RQ1.1 are presented in Table~\ref{tab:ProjectsInfectedbyMAS}   and Table~\ref{tab:GloballyRQ1}. We analyzed 15 combinations (CS-MAS pairs composed of 15 code smells and a Multiple Architectural Smell) for each of the 102 projects (1,530 analyses)  and for the data from all the projects merged together as a single project.  

Regarding the analysis performed with 111 projects separately, we decided not to consider the combinations (MAS-LC), (MAS-MC), and (MAS-SAK), due to the low number of infected projects (less than twelve). 
We found statistically significant results (p-value $<$0.05) for the remaining 13 combinations. 
However, only two combinations in two different projects showed a correlation higher than 0.5. 
For the other eleven combinations we found low correlations (tau $<$0.5) in a considerable number of projects (at least 32). 

According to our results, we found that only SC (Spaghetti Code) and LPL (Long Parameter List) have a low correlation with MAS.

Considering all the projects merged together, the results did not change and we found for the majority of the cases that the results were not statistically significant. 

The results confirm our Hypothesis 0, since the presence of a Multiple Architectural Smell does not depend on the presence of code smells in the 102 analyzed projects.

\subsection{RQ2:Is the presence of an architectural smell  independent from the presence of a \emph{category} of code smells?}
In order to answer this RQ, we considered the three categories of code smells reported in Section \ref{sec:background_categoryOfSmells}: \textit{Bloaters} (Bloat.), \textit{Dispensables }(Disp), and \textit{Object Orientation Abusers} (OOA). In this case, we considered all the code smells belonging to the same category as a single code smell. 

The results obtained for RQ2 are shown in Table~\ref{tab:ProjectsInfected2} and Table~\ref{tab:GloballyRQ2}. We analyzed nine combinations of AS-CS (pairs composed of three categories of code smells and three architectural smells) for each of the 102 projects (918 analyses) and for the data from all the projects merged together as a single project.

Regarding the analysis performed with 111 projects separately,  we found statistically significant results (p-value $<$0.05) for all the combinations. However, only two combinations with the same category of code smells (OOA) showed a correlation higher than 0.5 in two different projects, as shown in Table~\ref{tab:ProjectsInfected2}. 
For the other seven combinations, we found low correlations (tau $<$0.5) in a huge number of projects (at least 92). 
The results are similar to the one reported for the analysis of the non-categorized code smells in (Table~\ref{tab:ProjectsInfectedbyCD}, Table~\ref{tab:ProjectsInfectedbyHD}, and Table~\ref{tab:ProjectsInfectedbyUD}). 

According to our results, the most interesting low correlations we found are between the category of Object-Oriented Abuser (OOA) smells and the two HD and UD architectural smells.

Considering all the projects merged together, the results did not change and we found for the majority of the cases that the results were not statistically significant.

We can accept our Hypothesis 0, since the presence of an architectural smell does not depend on the presence of a \emph{category} of  code smells. Even though three projects were infected by the same category of code smells, we were unable to consider the results since the sample was too small.

\subsection{RQ2.1:Is the presence of a Multiple Architectural Smell independent from the presence of a \emph{category} of code smells?}
In order to answer this RQ, we considered the same category of code smells adopted in RQ2.
The results obtained for RQ2.1 are shown in Table~\ref{tab:ProjectsInfected2}  and Table~\ref{tab:GloballyRQ2}. We analyzed three combinations (CS-AS pairs composed of three categories of code smells and one Multiple Architectural Smell) for each of the 102 projects (306 analyses) and for the data from all the projects merged together as a single project. 

Regarding the analysis performed with 111 projects separately, we found statistically significant results (p-value $<$0.05) for all the combinations. 
However, only one combination showed a correlation higher than 0.5, and only in one project. 
For the other three combinations, we found low correlations (tau $<$0.5) in a huge number of projects (at least 98). 

According to our results, the most interesting low correlations we found are between the category the Bloater category and the MAS architectural smell. 

We can accept Hypothesis 0, since the presence of a Multiple Architectural Smell does not depend on the presence of a \emph{category} of code smells. 

Considering all the projects merged together, the results did not change and we found for the majority of the cases that the results were not statistically significant.
In conclusion we found very low correlations.  
Correlations in all the projects are very low, and merging the data did not help to increase the number of correlations.

\subsection{Lessons Learned} 
  \textit{Lesson Learned 1:} An architectural smell or Multiple Architectural Smells do not depend on code smells. As we can see from the analysis, statistically significant results were found in all projects for all cases of AS-CS pairs, and for MAS-LPL (Multiple Architectural Smell-Long Parameters List) or MAS-SC (Spaghetti Code). However, some code smells were found to infect projects more frequently than others and therefore the results of the analyses are more reliable for them. Considering all the analyses (4,590 for RQ1 and 1,530 for RQ1.1), 58.8\% of them provided statistically significant results and only 0.03\% of them showed a correlation higher than 0.5. 

Therefore, the main lesson learned from the analysis of these RQs is that architectural smells do not depend on code smells and therefore the refactoring of code smells  does not decrease the chances of removing  architectural smells. Moreover, the removal of an architectural smell  does not imply the removal or reduction of code smells, either. 
Hence, developers can  focus their attention on the refactoring of the more dangerous architectural smells.

\textit{Lesson Learned 2:} An architectural smell or Multiple Architectural Smells do not depend on categories of code smells. In this case, too, 78\% of them provided statistically significant results (918 for RQ2 and 306 for RQ 2.1), and only 0.3\% of them showed a correlation higher than 0.5. 

Even if we considered categories of code smells or some smells together, the results do not change, which confirms the need to analyze and remove smells at both levels, i.e., both code and architectural smells.

Oizumi et al\cite{Oizumi2016} outlined that design problems, structures that violates fundamental design principles, can be located by not considering only syntactical agglomeration of code smells, but also the semantically ones.
We have not considered semantically relations, but we  considered categories of smells  that can be viewed as a kind of semantically agglomeration. According to the categories  we found correlations between the category of Object-Oriented Abuser (OOA) smells and the two HD and UD architectural smells and between the Bloater category of smells and the MAS architectural smell.
Among the design problems considered in the study of  Oizumi et al and our AS,  only Cyclic Dependency problem is in common and according to the code smells only Long Method and Long Parameter List is in common, hence the results obtained in the two study are quite different and a next future development is related to consider the detection of other smells, that could be more relevant according to their impact on architectural problems . Moreover, we could remove from the code smells list of Section 3.1 those that could be seen more as design/architectural problems than problems at the code level, such as for example Swiss Army Knife and Spaghetti Code smells.
Moreover, in the Oizumi paper the design problems have been identified through the developers feedbacks by creating a “ground truth” of design problems  for the  set of considered projects. We detected the architectural problems/smells through the Arcan tool.
In conclusion, with respect to the results obtained we can observe that the relation between Spaghetti Code and all the other AS can be an expected relation. In any case all  the correlations have been found in few projects.

The independence between issues at the code level detected using the Technical Debt Index provided by SonarQube and those  at the architectural
level detected using the Architectural Debt Index computed by Arcan has been evaluated in a previous study \cite{Roveda2018}. 
Developers have to take care of
both possible sources of debt, as removing code debt does not necessarily imply the reduction/removal of architectural debt and vice versa.
The results described in this paper are in line with this previous result.

\textit{Lesson Learned 3:} Code smells and architectural smells correlations have to be further investigated as we will outline in future developments.
According to the results we found, we could expect to find a correlation between Spaghetti Code and  all the AS, as well as Spaghetti Code and Long Parameter List with MAS. We have not find correlations between some code smells and architectural smells previously found in the literature, since the tools we used were not able to detect some of these smells, such as for example Feature Envy and Divergent Change CS  and Scattered Functionality AS. Hence, it’s difficult to compare our results with previous results of the literature.
For example in the work of Oizumi \cite{Oizumi2016} other AS (design problems) have been considered with the exception of Cyclic Dependency. They found that some code anomalies often flock together in order to embody a design problem.  They analyzed not only syntactic agglomerations of code anomalies, but also the semantic ones, as possible indicators of design problems. We have not considered in our work these kinds of agglomerations, but we considered some categories of code smells. Moreover, in their work design problems have been identified through the developers feedbacks by creating a “ground truth” of design problems  for the  set of considered projects. While we  detected the architectural problems/smells through an automated  support .
Hence, further investigations have to be done and this study can also be used to better design future analysis.

\section{Threats to Validity}
\label{sec:Threats}
In this Section, we introduce the threats to validity, following the structure suggested by Yin~\cite{YinCaseStudies2009}, reporting construct validity, internal validity, external validity, and reliability. Moreover, we will also discuss the different tactics adopted to mitigate them.

\subsection{Construct  Validity}
\textit{Construct Validity} concerns the identification of the measures adopted for the concepts studied in this work.
We used two widely accepted tools to measure code and architectural smells, but we are aware that other tools could have reported different results or could have detected other types of smells. 
To reduce the threat related to the data analysis technique adopted,  we used  Kendall rank correlation, since it has less gross error sensitivity enabling a more robust analysis with a smaller asymptotic variance~\cite{Croux2010}.

\subsection{Internal Validity}
 
Threats to \textit{Internal Validity} concern factors that might have influenced the results obtained. 
Regarding this threat, the main issue is related to the detection accuracy of the adopted tools. 
For this purpose, we relied on existing detection tools already adopted in previous research studies.  
Regarding code smell detection, we relied on the DECOR rules. We would like to point out that the SonarQube ''Antipatterns-CodeSmell'' plugin adopts the exact rules defined by Moha et al.~\cite{Moha2007a}. We are aware that the results could be influenced by the presence of false negatives and positives. For this reason,  Moha et al. reported  a precision higher than 60\% for DECOR and a recall of 100\% on a selected set of projects. Moreover, in our previous work~\cite{TaibiIST2017}, two authors independently manually validated a subset of smell instances, reporting a mean precision of 78\%. The results of the validation analyzed in~\cite{TaibiIST2017} are also available in its replication package~\cite{TaibiRawData2017}. 

The evaluation of Arcan's detection performance in two  industrial case studies based on the feedback of the developers is described in~\cite{ArcelliFontana2017}, where the authors report a precision of 100\%, since Arcan found only correct instances of architectural smells,  and a recall of 66\%.  The developers reported five more architectural smells, which were false negatives related to 180 external components outside the tool's scope of analysis.
According to the recall value, some AS can be missed and therefore, we might have failed to detect some correlations.
Moreover, the manual validation of the  Arcan's detection results has been done on ten open source projects d~\cite{ArcelliFontana2016c} and  a multiple case study on several architectural
smells detected by Arcan has been conducted  on  four industrial projects ~\cite{Martini2018}, with the aim to evaluate the negative
impact of  the architectural smells based on the feedbacks from practitioners.  
According to this study  practitioners appreciated the support of the automatic detection of AS provided by Arcan~\cite{Martini2018}.


Based on the previous assumptions, the presence of possible false positives and false negatives is mitigated also by the large sample of analyzed projects and by the  high precision and recall values of the results of the two detection tools.

\subsection{External Validity}
Threats to \textit{External Validity} concern the generalization of the results obtained. We cannot claim that our results fully represent every Java project. In order to mitigate this issue, we considered a large set of projects with different characteristics, in particular a set of 102 well-known Java projects included in the Qualitas Corpus data set. This data set includes projects from different domains, of different sizes, and with different architectures. 
Hence, this data set is representative and useful for reducing the possibility that the results might not be generalizable.
As for the selection of the projects for this study, the adoption of Open Source projects instead of commercial ones,  should not have influenced the results of this work. Open source projects are now considered at the same level of quality of closed source projects~\cite{LenarduzziOSS2019}. Therefore,  we hypothesized that commercial projects, in similar domains,  would have reported a similar result.
We have analyzed only Java projects, hence we can not generalize our results to projects in different programming languages.

We used two available tools to measure code and architectural smells. Not all of the defined code smells and architectural smells in the literature  are detected by these tools.
Hence, we cannot claim that our results will hold for any code smell or architectural smell.  In order to mitigate this issue, we considered a large number of code smells, but other code smells can be considered in future work, such as for example the Feature Envy smell \cite{Fowler1999}.
The detection of this smell could have a significant impact on the results:  when envying classes, or in the case of classes that are being envied and very scattered in a software project, this might actually represent an indicator of a design or architecture-level problem. 
While for architectural smells, we have considered only four architectural smells, since the availability of tools able to detect several architectural smells \footnote{at least at the time when we performed this study.} is reduced with respect to code smells
Hence, we have to extend this study by considering other architectural smells and, in particular, other smells not focused only on dependency issues. To accomplish this task, we have to extend Arcan with the ability to detect such new architectural smells or use another available tool.

Moreover, to enable the replicability of this work, we provided a complete replication package~\cite{TaibiRawData2017}.



\subsection{Reliability}
Threats to \textit{Reliability} refer to the correctness of the conclusions reached in the study. We applied non-parametric tests and rank-based correlation methods, as software metrics often do not have normal distributions. 
We used a standard R package to perform all statistical analyses since it allows simple replications and gives good confidence on the quality of the results.

\section{Conclusion and Future Development}
\label{sec:Conclusions and Future Developments}
In this work, we conducted a large-scale empirical study investigating the correlations between code smells and architectural smells.
We detected code smells and architectural smells in 102 Java projects of the Qualitas Corpus data set~\cite{qualitas.class} by means of two smell detection tools, the SonarQube ''Antipatterns-CodeSmells plugin'' for code smells and Arcan for architectural smells.

We found empirical evidence on the independence between code smells and architectural smells.
Therefore, we can assume that the presence of  code smells does not imply the presence of  architectural smells and vice versa. 
 This result can be useful for developers, since they  cannot focus only on the refactoring of code smells, but also need to pay particular attention to the more dangerous architectural smells.
 Moreover, this result can stimulate research in this direction  to enhance the detection of architectural issues such as architectural smells. Also, it may provide an incentive for studying and providing support for the automated refactoring of AS.

Future work will include the replication of this work considering different projects and their historical evolution. In particular, we would like to consider projects in different categories and evaluate whether the domain of the project might have an impact on this study.
We have to extend our study by considering  other code and  architectural smells.
Regarding code smells, we have to consider at least the Feature Envy and Divergent Change smells,  not detected by the tool we used. Hence, we could revisit the classification of CS in the different categories according to the new introduced CS.
Regarding architectural smells, we have to consider a larger set of smells  not focused only on dependency issues, such as those  identified, for example, by Macia et al.\cite{Macia2012b} and Garcia et al.\cite{Garcia2009}.
To accomplish this task, we could extend Arcan  to detect these new architectural smells or exploit any other tool  available in the future.
We are also working  on the extension of Arcan to detect architectural smells in microservice architectures~\cite{TaibiIEEESW18}.

Research interest on architectural smells is increasing, and this will certainly lead to the development of new tools or the extension of the existing ones. 

Therefore, we believe there is a need for more empirical investigations in this domain, so as to understand whether (a)  the presence of other code smells implies the presence of one or more architectural smells; (b) the independence between architectural smells and code smells is still true  when considering other architectural smells not currently detected by Arcan or by other available tools; (c) the results obtained are valid for other projects in different domains.

Moreover, as outlined  by Kouroshfar et al.~\cite{Kouroshfar:2015}, to improve the accuracy  of bug prediction one should also take the software architecture of the project into consideration. Hence, in the near future we would like to study potential correlations between architectural smells and bugs as well as potential correlations with other issues detected through SonarQube~\cite{Saarimaki2019}. 

\section*{References}

\bibliographystyle{model1-num-names}
\bibliography{manuscript}
\appendix
\section{The analyzed projects}
\label{sec:appendixa}

\begin{landscape}
{\scriptsize
\centering
\setlength{\tabcolsep}{4pt}
\begin{longtable}[c]{@{}l|llll|llll|llllllllllllll@{}}
\caption{Number of Architectural Smells, Category of Code Smells, and Code Smells infecting the analyzed projects} 
\label{tab:ProjectsInfectedDetails}\\
\hline
\multirow {2}{*}{Project} & \multicolumn{4}{c|}{Architectural Smells} & \multicolumn{4}{c|}{Category of Code Smells} & \multicolumn{13}{c}{Code Smells}\\ \cline{2-23}
	&	\rotatebox{90}{UD}	&	\rotatebox{90}{HL}	&	\rotatebox{90}{CD}	&	\rotatebox{90}{MAS}	&	\rotatebox{90}{Bloat.}	&	\rotatebox{90}{Disp.}	&	\rotatebox{90}{Enc.}	&	\rotatebox{90}{OOA}	&	\rotatebox{90}{AS}	&	\rotatebox{90}{BCSA}	&	\rotatebox{90}{DsP}	&	\rotatebox{90}{CC}	&	\rotatebox{90}{DC}	&	\rotatebox{90}{LC}	&	\rotatebox{90}{LzC}	&	\rotatebox{90}{LM}	&	\rotatebox{90}{LPL}	&	\rotatebox{90}{MFnC}	&	\rotatebox{90}{MC}	&	\rotatebox{90}{RPB}	&	\rotatebox{90}{SC}	&	\rotatebox{90}{SG} \\ 
\hline
\endfirsthead 

\multicolumn{23}{l}
{{\bfseries \tablename\ \thetable{} -- continued from previous page}} \\
\hline 
\multirow {2}{*}{Project} & \multicolumn{4}{c|}{Architectural Smells} & \multicolumn{4}{c|}{Category of Code Smells} & \multicolumn{13}{c}{Code Smells}\\ \cline{2-23}
	&	\rotatebox{90}{UD}	&	\rotatebox{90}{HL}	&	\rotatebox{90}{CD}	&	\rotatebox{90}{MAS}	&	\rotatebox{90}{Bloat.}	&	\rotatebox{90}{Disp.}	&	\rotatebox{90}{Enc.}	&	\rotatebox{90}{OOA}	&	\rotatebox{90}{ASG}	&	\rotatebox{90}{BCSA}	&	\rotatebox{90}{DsP}	&	\rotatebox{90}{CC}	&	\rotatebox{90}{DC}	&	\rotatebox{90}{LC}	&	\rotatebox{90}{LzC}	&	\rotatebox{90}{LM}	&	\rotatebox{90}{LPL}	&	\rotatebox{90}{MFnC}	&	\rotatebox{90}{MC}	&	\rotatebox{90}{RPB}	&	\rotatebox{90}{SC}	&	\rotatebox{90}{SG} \\ 
\hline
\endhead

\multicolumn{23}{l}{{Continued on next page}} \\ 
\endfoot

\hline 
\endlastfoot 
aoi	&	6	&	7	&	11110	&	11123	&	225	&	191	&	42	&	31	&	31	&	0	&	42	&	78	&	188	&	0	&	3	&	108	&	39	&	0	&	0	&	0	&	3	&	0	\\
argouml	&	3	&	25	&	1833	&	1861	&	854	&	1344	&	62	&	17	&	17	&	3	&	61	&	333	&	1294	&	0	&	50	&	429	&	89	&	3	&	1	&	0	&	6	&	2	\\
aspectj	&	8	&	52	&	48064	&	48124	&	974	&	1564	&	353	&	93	&	93	&	5	&	353	&	437	&	1501	&	0	&	63	&	352	&	183	&	2	&	0	&	6	&	10	&	2	\\
axion	&	1	&	5	&	158	&	164	&	85	&	40	&	8	&	3	&	3	&	0	&	8	&	31	&	39	&	0	&	1	&	46	&	7	&	1	&	0	&	0	&	0	&	0	\\
azureus	&	6	&	168	&	162357	&	162531	&	1193	&	704	&	173	&	139	&	139	&	65	&	111	&	428	&	494	&	0	&	210	&	462	&	303	&	0	&	62	&	139	&	19	&	25	\\
cjdbc	&	7	&	35	&	1632	&	1674	&	365	&	136	&	21	&	54	&	54	&	0	&	21	&	140	&	132	&	0	&	4	&	165	&	60	&	0	&	0	&	0	&	6	&	0	\\
castor	&	9	&	58	&	1510	&	1577	&	1511	&	1017	&	32	&	11	&	11	&	4	&	32	&	511	&	956	&	0	&	61	&	573	&	427	&	0	&	0	&	4	&	4	&	1	\\
cayenne	&	2	&	1	&	12	&	15	&	1241	&	552	&	18	&	25	&	25	&	4	&	18	&	479	&	343	&	0	&	209	&	655	&	106	&	1	&	0	&	32	&	0	&	25	\\
checkstyle	&	4	&	3	&	253	&	260	&	446	&	50	&	10	&	5	&	5	&	2	&	10	&	186	&	50	&	0	&	0	&	158	&	100	&	2	&	0	&	2	&	0	&	0	\\
cobertura	&	6	&	4	&	38	&	48	&	78	&	49	&	26	&	3	&	3	&	2	&	26	&	30	&	48	&	0	&	1	&	31	&	17	&	0	&	0	&	0	&	0	&	0	\\
collections	&	4	&	5	&	360	&	369	&	241	&	411	&	2	&	2	&	2	&	1	&	2	&	77	&	409	&	0	&	2	&	107	&	57	&	0	&	0	&	0	&	0	&	4	\\
colt	&	6	&	9	&	885	&	900	&	85	&	109	&	6	&	22	&	22	&	1	&	6	&	17	&	95	&	0	&	14	&	50	&	18	&	0	&	0	&	2	&	5	&	0	\\
columba	&	5	&	60	&	2294	&	2359	&	623	&	146	&	24	&	29	&	29	&	0	&	24	&	169	&	134	&	0	&	12	&	253	&	201	&	0	&	0	&	0	&	0	&	0	\\
compiere	&	9	&	19	&	8282	&	8310	&	1351	&	1292	&	81	&	734	&	734	&	2	&	81	&	412	&	1265	&	2	&	27	&	516	&	415	&	6	&	0	&	1	&	5	&	1	\\
derby	&	4	&	51	&	9648	&	9703	&	1455	&	739	&	163	&	98	&	98	&	11	&	163	&	542	&	581	&	0	&	158	&	604	&	303	&	6	&	0	&	9	&	16	&	6	\\
displaytag	&	1	&	5	&	117	&	123	&	197	&	81	&	0	&	2	&	2	&	2	&	0	&	61	&	81	&	0	&	0	&	79	&	57	&	0	&	0	&	1	&	0	&	0	\\
drawswf	&	2	&	14	&	364	&	380	&	102	&	57	&	7	&	12	&	12	&	10	&	7	&	40	&	55	&	0	&	2	&	53	&	9	&	0	&	0	&	10	&	0	&	1	\\
emma	&	3	&	8	&	187	&	198	&	79	&	41	&	19	&	0	&	0	&	0	&	19	&	21	&	39	&	0	&	2	&	28	&	30	&	0	&	0	&	0	&	0	&	3	\\
exoportal	&	0	&	49	&	370	&	419	&	998	&	269	&	44	&	77	&	77	&	11	&	44	&	263	&	228	&	0	&	41	&	357	&	377	&	1	&	0	&	8	&	7	&	3	\\
findbugs	&	10	&	13	&	9111	&	9134	&	449	&	97	&	14	&	34	&	34	&	0	&	14	&	174	&	72	&	0	&	25	&	202	&	72	&	1	&	0	&	0	&	6	&	0	\\
fitjava	&	2	&	0	&	32	&	34	&	32	&	77	&	31	&	15	&	15	&	1	&	31	&	9	&	75	&	0	&	2	&	15	&	8	&	0	&	0	&	0	&	0	&	1	\\
fitlibraryforfitnesse	&	6	&	38	&	2436	&	2480	&	607	&	161	&	75	&	27	&	27	&	11	&	75	&	152	&	102	&	0	&	59	&	229	&	224	&	2	&	0	&	4	&	3	&	0	\\
freecol	&	15	&	20	&	41088	&	41123	&	323	&	86	&	6	&	6	&	6	&	0	&	6	&	123	&	81	&	0	&	5	&	159	&	41	&	0	&	0	&	0	&	0	&	0	\\
freecs	&	4	&	6	&	742	&	752	&	78	&	66	&	26	&	13	&	13	&	0	&	26	&	27	&	66	&	0	&	0	&	29	&	21	&	1	&	0	&	0	&	0	&	0	\\
freemind	&	9	&	16	&	4350	&	4375	&	181	&	54	&	48	&	27	&	27	&	10	&	11	&	64	&	43	&	0	&	11	&	89	&	28	&	0	&	37	&	34	&	6	&	3	\\
galleon	&	8	&	6	&	1788	&	1802	&	153	&	125	&	18	&	20	&	20	&	0	&	18	&	66	&	125	&	0	&	0	&	50	&	37	&	0	&	0	&	0	&	3	&	0	\\
ganttproject	&	5	&	20	&	1852	&	1877	&	222	&	51	&	16	&	13	&	13	&	0	&	16	&	79	&	39	&	0	&	12	&	103	&	39	&	1	&	0	&	0	&	3	&	0	\\
hadoop	&	9	&	48	&	6865	&	6922	&	994	&	789	&	46	&	103	&	103	&	11	&	46	&	326	&	753	&	0	&	36	&	550	&	117	&	1	&	0	&	10	&	11	&	6	\\
heritrix	&	6	&	16	&	996	&	1018	&	261	&	63	&	20	&	37	&	37	&	0	&	20	&	89	&	61	&	0	&	2	&	138	&	34	&	0	&	0	&	0	&	1	&	0	\\
hsqldb	&	9	&	11	&	10606	&	10626	&	257	&	193	&	44	&	42	&	42	&	3	&	44	&	112	&	162	&	1	&	31	&	109	&	35	&	0	&	0	&	1	&	8	&	0	\\
htmlunit	&	2	&	7	&	10408	&	10417	&	357	&	118	&	0	&	0	&	0	&	0	&	0	&	176	&	117	&	0	&	1	&	178	&	3	&	0	&	0	&	0	&	0	&	0	\\
informa	&	3	&	3	&	49	&	55	&	65	&	42	&	0	&	8	&	8	&	4	&	0	&	24	&	42	&	0	&	0	&	31	&	10	&	0	&	0	&	3	&	0	&	0	\\
ireport	&	9	&	44	&	14344	&	14397	&	893	&	838	&	36	&	39	&	39	&	1	&	36	&	303	&	829	&	2	&	9	&	270	&	318	&	0	&	0	&	1	&	8	&	0	\\
itext	&	6	&	6	&	3519	&	3531	&	215	&	89	&	18	&	17	&	17	&	0	&	18	&	113	&	63	&	1	&	26	&	73	&	28	&	0	&	0	&	0	&	3	&	0	\\
ivatagroupware	&	3	&	20	&	14	&	37	&	105	&	25	&	2	&	1	&	1	&	0	&	2	&	38	&	19	&	0	&	6	&	44	&	23	&	0	&	0	&	0	&	0	&	0	\\
jag	&	1	&	7	&	422	&	430	&	72	&	29	&	13	&	11	&	11	&	0	&	13	&	19	&	20	&	0	&	9	&	30	&	23	&	0	&	0	&	0	&	4	&	0	\\
james	&	3	&	6	&	100	&	109	&	144	&	67	&	2	&	1	&	1	&	0	&	2	&	56	&	60	&	0	&	7	&	81	&	7	&	0	&	0	&	1	&	0	&	1	\\
jasml	&	0	&	0	&	0	&	0	&	21	&	5	&	27	&	3	&	3	&	0	&	27	&	10	&	3	&	0	&	2	&	4	&	5	&	2	&	0	&	0	&	0	&	0	\\
jasperreports	&	7	&	21	&	1829	&	1857	&	820	&	306	&	0	&	3	&	3	&	1	&	0	&	285	&	300	&	0	&	6	&	282	&	253	&	0	&	0	&	0	&	1	&	0	\\
javacc	&	4	&	2	&	47	&	53	&	70	&	57	&	38	&	23	&	23	&	2	&	38	&	30	&	54	&	0	&	3	&	28	&	11	&	1	&	0	&	0	&	9	&	0	\\
jboss	&	9	&	93	&	2658	&	2760	&	3364	&	1501	&	305	&	358	&	358	&	17	&	305	&	914	&	1330	&	0	&	171	&	1251	&	1197	&	2	&	0	&	11	&	40	&	3	\\
jchempaint	&	5	&	37	&	1737	&	1779	&	914	&	823	&	38	&	44	&	44	&	4	&	38	&	386	&	693	&	0	&	130	&	426	&	102	&	0	&	0	&	1	&	8	&	0	\\
jedit	&	8	&	14	&	11509	&	11531	&	257	&	138	&	47	&	22	&	22	&	1	&	47	&	104	&	89	&	0	&	49	&	131	&	20	&	2	&	0	&	1	&	0	&	0	\\
jena	&	6	&	21	&	6257	&	6284	&	324	&	117	&	71	&	115	&	115	&	28	&	66	&	134	&	73	&	0	&	44	&	153	&	36	&	1	&	5	&	71	&	13	&	7	\\
jext	&	6	&	13	&	1255	&	1274	&	277	&	153	&	27	&	16	&	16	&	1	&	27	&	124	&	150	&	0	&	3	&	123	&	29	&	1	&	0	&	1	&	2	&	0	\\
jFin DateMath	&	0	&	2	&	0	&	2	&	48	&	20	&	3	&	3	&	3	&	0	&	3	&	19	&	18	&	0	&	2	&	25	&	4	&	0	&	0	&	0	&	0	&	0	\\
jfreechart	&	9	&	15	&	245	&	269	&	440	&	331	&	5	&	29	&	29	&	0	&	5	&	134	&	326	&	0	&	5	&	214	&	92	&	0	&	0	&	0	&	1	&	0	\\
jgraph	&	6	&	8	&	908	&	922	&	81	&	67	&	37	&	44	&	44	&	0	&	37	&	26	&	56	&	0	&	11	&	44	&	11	&	0	&	0	&	0	&	0	&	0	\\
jgraphpad	&	4	&	4	&	186	&	194	&	166	&	35	&	21	&	32	&	32	&	0	&	21	&	56	&	35	&	0	&	0	&	64	&	46	&	0	&	0	&	1	&	3	&	1	\\
jgrapht	&	1	&	6	&	58	&	65	&	96	&	19	&	8	&	2	&	2	&	0	&	8	&	33	&	16	&	0	&	3	&	58	&	5	&	0	&	0	&	0	&	0	&	0	\\
jgroups	&	4	&	7	&	859	&	870	&	301	&	144	&	13	&	24	&	24	&	1	&	13	&	113	&	135	&	1	&	9	&	169	&	18	&	0	&	0	&	1	&	1	&	0	\\
jhotdraw	&	8	&	22	&	827	&	857	&	356	&	274	&	9	&	8	&	8	&	4	&	9	&	140	&	241	&	0	&	33	&	146	&	70	&	0	&	0	&	0	&	0	&	0	\\
jmeter	&	7	&	35	&	4464	&	4506	&	498	&	222	&	1	&	0	&	0	&	0	&	1	&	117	&	170	&	0	&	52	&	196	&	185	&	0	&	0	&	0	&	0	&	0	\\
jmoney	&	0	&	3	&	302	&	305	&	34	&	14	&	0	&	4	&	4	&	1	&	0	&	10	&	7	&	0	&	7	&	14	&	10	&	0	&	0	&	0	&	0	&	0	\\
joggplayer	&	3	&	3	&	212	&	218	&	122	&	18	&	13	&	16	&	16	&	1	&	13	&	41	&	16	&	0	&	2	&	44	&	37	&	0	&	0	&	0	&	5	&	0	\\
jparse	&	1	&	1	&	122	&	124	&	41	&	21	&	7	&	3	&	3	&	1	&	1	&	14	&	21	&	0	&	0	&	15	&	12	&	0	&	6	&	0	&	1	&	0	\\
jpf	&	1	&	2	&	42	&	45	&	35	&	8	&	0	&	4	&	4	&	0	&	0	&	11	&	7	&	0	&	1	&	19	&	5	&	0	&	0	&	0	&	0	&	0	\\
jrefactory	&	4	&	37	&	2901	&	2942	&	805	&	309	&	33	&	23	&	23	&	29	&	27	&	260	&	309	&	0	&	0	&	310	&	233	&	2	&	6	&	74	&	6	&	5	\\
jruby	&	12	&	46	&	147592	&	147650	&	636	&	208	&	68	&	45	&	45	&	9	&	68	&	304	&	179	&	0	&	29	&	272	&	58	&	2	&	0	&	4	&	9	&	6	\\
jspwiki	&	6	&	17	&	1610	&	1633	&	317	&	61	&	9	&	26	&	26	&	0	&	9	&	80	&	60	&	0	&	1	&	139	&	98	&	0	&	0	&	0	&	0	&	0	\\
jsXe	&	6	&	7	&	368	&	381	&	72	&	29	&	6	&	2	&	2	&	0	&	6	&	24	&	14	&	0	&	15	&	34	&	12	&	2	&	0	&	0	&	0	&	1	\\
jtopen	&	1	&	3	&	3690	&	3694	&	982	&	748	&	10	&	50	&	50	&	0	&	10	&	461	&	655	&	0	&	93	&	311	&	209	&	1	&	0	&	0	&	6	&	0	\\
jung	&	2	&	14	&	200	&	216	&	173	&	112	&	6	&	10	&	10	&	0	&	6	&	48	&	112	&	0	&	0	&	92	&	33	&	0	&	0	&	0	&	2	&	0	\\
junit	&	2	&	11	&	183	&	196	&	79	&	49	&	3	&	3	&	3	&	7	&	3	&	15	&	46	&	0	&	3	&	35	&	29	&	0	&	0	&	17	&	1	&	6	\\
log4j	&	3	&	13	&	411	&	427	&	190	&	91	&	3	&	11	&	11	&	2	&	3	&	61	&	89	&	0	&	2	&	107	&	22	&	0	&	0	&	0	&	2	&	1	\\
lucene	&	2	&	66	&	3786	&	3854	&	608	&	364	&	44	&	26	&	26	&	3	&	44	&	190	&	353	&	0	&	11	&	284	&	134	&	0	&	0	&	0	&	2	&	4	\\
marauroa	&	5	&	12	&	264	&	281	&	94	&	30	&	6	&	0	&	0	&	0	&	6	&	26	&	28	&	0	&	2	&	57	&	11	&	0	&	0	&	0	&	0	&	0	\\
maven	&	6	&	27	&	691	&	724	&	331	&	99	&	1	&	0	&	0	&	5	&	1	&	115	&	80	&	0	&	19	&	149	&	67	&	0	&	0	&	5	&	0	&	1	\\
megamek	&	4	&	20	&	11532	&	11556	&	581	&	1076	&	61	&	24	&	24	&	1	&	61	&	387	&	1076	&	0	&	0	&	182	&	7	&	5	&	0	&	1	&	6	&	0	\\
mvnforum	&	3	&	26	&	1071	&	1100	&	331	&	294	&	10	&	18	&	18	&	1	&	10	&	118	&	240	&	0	&	54	&	128	&	85	&	0	&	0	&	1	&	2	&	0	\\
myfaces core	&	7	&	41	&	1267	&	1315	&	924	&	1849	&	5	&	5	&	5	&	16	&	5	&	294	&	1830	&	0	&	19	&	299	&	331	&	0	&	0	&	53	&	0	&	35	\\
nakedobjects	&	2	&	99	&	2578	&	2679	&	1082	&	322	&	44	&	34	&	34	&	0	&	44	&	398	&	223	&	0	&	99	&	557	&	127	&	0	&	0	&	0	&	2	&	1	\\
nekohtml	&	1	&	2	&	15	&	18	&	13	&	15	&	0	&	1	&	1	&	0	&	0	&	6	&	6	&	0	&	9	&	7	&	0	&	0	&	0	&	0	&	0	&	0	\\
openjms	&	2	&	19	&	301	&	322	&	230	&	60	&	0	&	3	&	3	&	0	&	0	&	64	&	46	&	0	&	14	&	127	&	39	&	0	&	0	&	0	&	0	&	0	\\
oscache	&	1	&	4	&	45	&	50	&	63	&	8	&	5	&	9	&	9	&	0	&	5	&	16	&	8	&	0	&	0	&	25	&	22	&	0	&	0	&	0	&	1	&	0	\\
picocontainer	&	1	&	4	&	129	&	134	&	53	&	11	&	0	&	0	&	0	&	1	&	0	&	21	&	7	&	0	&	4	&	32	&	0	&	0	&	0	&	18	&	0	&	1	\\
pmd	&	2	&	17	&	310	&	329	&	379	&	77	&	18	&	5	&	5	&	4	&	18	&	120	&	52	&	0	&	25	&	167	&	92	&	0	&	0	&	2	&	0	&	2	\\
poi	&	13	&	43	&	3491	&	3547	&	956	&	306	&	25	&	20	&	20	&	7	&	25	&	408	&	270	&	0	&	36	&	432	&	113	&	3	&	0	&	0	&	1	&	2	\\
pooka	&	7	&	8	&	10070	&	10085	&	113	&	93	&	64	&	48	&	48	&	22	&	53	&	29	&	75	&	0	&	18	&	74	&	10	&	0	&	11	&	12	&	5	&	1	\\
proguard	&	2	&	15	&	287	&	304	&	278	&	34	&	73	&	1	&	1	&	0	&	73	&	118	&	34	&	0	&	0	&	124	&	35	&	1	&	0	&	0	&	1	&	0	\\
quartz	&	1	&	8	&	198	&	207	&	93	&	48	&	0	&	0	&	0	&	0	&	0	&	31	&	44	&	0	&	4	&	45	&	17	&	0	&	0	&	0	&	0	&	0	\\
quickserver	&	1	&	6	&	288	&	295	&	89	&	57	&	13	&	9	&	9	&	0	&	13	&	29	&	57	&	0	&	0	&	38	&	22	&	0	&	0	&	0	&	1	&	0	\\
quilt	&	0	&	3	&	70	&	73	&	48	&	10	&	7	&	6	&	6	&	0	&	7	&	8	&	10	&	0	&	0	&	23	&	17	&	0	&	0	&	0	&	0	&	0	\\
roller	&	6	&	23	&	371	&	400	&	402	&	190	&	45	&	54	&	54	&	0	&	45	&	96	&	189	&	0	&	1	&	165	&	141	&	0	&	0	&	0	&	5	&	0	\\
rssowl	&	10	&	51	&	17009	&	17070	&	254	&	192	&	37	&	39	&	39	&	0	&	37	&	95	&	111	&	0	&	81	&	110	&	29	&	20	&	0	&	0	&	2	&	0	\\
sablecc	&	1	&	1	&	44	&	46	&	92	&	80	&	11	&	4	&	4	&	0	&	11	&	25	&	49	&	0	&	31	&	30	&	37	&	0	&	0	&	0	&	3	&	0	\\
sandmark	&	2	&	23	&	760	&	785	&	403	&	185	&	76	&	37	&	37	&	14	&	76	&	149	&	168	&	0	&	17	&	185	&	69	&	0	&	0	&	24	&	2	&	2	\\
springframework	&	7	&	100	&	3604	&	3711	&	2075	&	815	&	36	&	27	&	27	&	2	&	36	&	614	&	719	&	0	&	96	&	777	&	683	&	1	&	0	&	1	&	0	&	0	\\
squirrelsql	&	0	&	2	&	231	&	233	&	19	&	1	&	0	&	0	&	0	&	0	&	0	&	6	&	0	&	0	&	1	&	12	&	1	&	0	&	0	&	0	&	0	&	0	\\
struts	&	5	&	43	&	1241	&	1289	&	790	&	431	&	30	&	30	&	30	&	4	&	30	&	331	&	397	&	0	&	34	&	433	&	26	&	0	&	0	&	3	&	7	&	1	\\
sunflow	&	3	&	7	&	850	&	860	&	94	&	33	&	4	&	0	&	0	&	0	&	4	&	35	&	33	&	0	&	0	&	37	&	22	&	0	&	0	&	0	&	0	&	0	\\
tapestry	&	7	&	29	&	971	&	1007	&	745	&	49	&	5	&	5	&	5	&	0	&	5	&	245	&	49	&	0	&	0	&	395	&	105	&	0	&	0	&	0	&	0	&	0	\\
tomcat	&	5	&	35	&	1891	&	1931	&	612	&	909	&	22	&	68	&	68	&	3	&	22	&	229	&	795	&	0	&	114	&	263	&	119	&	1	&	0	&	3	&	10	&	1	\\
trove	&	0	&	0	&	17	&	17	&	22	&	8	&	0	&	0	&	0	&	0	&	0	&	8	&	5	&	0	&	3	&	9	&	5	&	0	&	0	&	0	&	0	&	0	\\
velocity	&	2	&	14	&	287	&	303	&	172	&	53	&	11	&	4	&	4	&	0	&	11	&	64	&	52	&	0	&	1	&	92	&	15	&	1	&	0	&	0	&	0	&	0	\\
wct	&	4	&	35	&	685	&	724	&	259	&	162	&	8	&	7	&	7	&	0	&	8	&	99	&	126	&	0	&	36	&	141	&	19	&	0	&	0	&	0	&	0	&	0	\\
webmail	&	4	&	7	&	117	&	128	&	50	&	29	&	2	&	4	&	4	&	1	&	2	&	17	&	27	&	0	&	2	&	29	&	4	&	0	&	0	&	2	&	0	&	0	\\
weka	&	4	&	32	&	5179	&	5215	&	654	&	639	&	56	&	85	&	85	&	0	&	56	&	206	&	578	&	0	&	61	&	334	&	114	&	0	&	0	&	0	&	4	&	0	\\
xalan	&	4	&	21	&	8027	&	8052	&	567	&	439	&	33	&	7	&	7	&	10	&	33	&	223	&	313	&	0	&	126	&	228	&	115	&	1	&	0	&	8	&	1	&	5	\\
xerces	&	2	&	12	&	1130	&	1144	&	367	&	439	&	31	&	13	&	13	&	5	&	31	&	116	&	409	&	0	&	30	&	104	&	146	&	1	&	0	&	0	&	2	&	0	\\
xmojo	&	0	&	1	&	4	&	5	&	10	&	16	&	3	&	2	&	2	&	0	&	3	&	3	&	14	&	0	&	2	&	6	&	1	&	0	&	0	&	0	&	1	&	0	\\
antlr	&	4	&	8	&	601	&	613	&	1	&	0	&	0	&	0	&	0	&	0	&	0	&	1	&	0	&	0	&	0	&	0	&	0	&	0	&	0	&	0	&	0	&	0	\\
ant	&	5	&	13	&	2511	&	2529	&	551	&	152	&	11	&	3	&	3	&	24	&	8	&	135	&	102	&	0	&	50	&	213	&	203	&	0	&	3	&	74	&	1	&	3	\\

\hline
\end{longtable}
}
\end{landscape}

\end{document}